\documentclass[preprint,aps,prd,showpacs,nofootinbib,superscriptaddress,floatfix]{revtex4}
\usepackage{graphicx}
\usepackage{amsmath}
\usepackage{amssymb}
\usepackage{bm}
\usepackage{bbm}
\begin{document}
\preprint{ANL-HEP-PR-07-79}

\title{\mbox{}\\[10pt]
Resummation of Relativistic Corrections to
$\bm{ e^{+}e^{-}\to J/\psi+\eta_{c}}$
}
%%%%%%%%%%%%%%%%%%%%%%%%%%%%%%%%%%%%%%%%%%%%%%%%%%%%%%%%%%%%%%%%%%%%%%%%%%%%%%
\author{Geoffrey T. Bodwin}
\affiliation{
High Energy Physics Division, Argonne National Laboratory,\\
9700 S. Cass Avenue, Argonne, Illinois 60439, USA}

\author{Jungil Lee}
\affiliation{
High Energy Physics Division, Argonne National Laboratory,\\
9700 S. Cass Avenue, Argonne, Illinois 60439, USA}
\affiliation{
Department of Physics, Korea University, Seoul 136-701, Korea}

\author{Chaehyun Yu}
\affiliation{
Department of Physics, Korea University, Seoul 136-701, Korea}

%\date{\today}
%%%%%%%%%%%%%%%%%%%%%%%%%%%%%%%%%%%%%%%%%%%%%%%%%%%%%%%%%%%%%%%%%%%%%%%%%%%%%%
\begin{abstract}
We present a new calculation, in the nonrelativistic QCD (NRQCD)
factorization formalism, of the relativistic corrections to the
double-charmonium cross section $\sigma[e^+e^-\to J/\psi+\eta_c]$ at the
energy of the Belle and BABAR experiments. In comparison with previous
work, our calculation contains several refinements. These include the
use of the improved results for the nonperturbative NRQCD matrix
elements, the resummation of a class of relativistic corrections, the
use of the vector-meson-dominance method to calculate the fragmentation
contribution to the pure QED amplitude, the inclusion of the effects of
the running of $\alpha$, and the inclusion of the contribution that
arises from the interference between the relativistic corrections and the
corrections of next-to-leading order in $\alpha_s$. We also present a
detailed estimate of the theoretical uncertainty. We
conclude that the discrepancy between the theoretical prediction for
$\sigma[e^+e^-\to J/\psi+\eta_c]$ and the experimental measurements has
been resolved.
\end{abstract}

% 12.38.-t  Quantum chromodynamics
% 12.39.St  Factorization
% 12.40.Vv  Vector-meson dominance 
% 13.20.Gd  Decays of J/psi, Upsilon, and other quarkonia (Leptonic decay)
% 13.25.Gv  Decays of J/psi, Upsilon, and other quarkonia (Hadronic decay)
% 13.66.Bc  Hadron production in e-e+ interactions  
% 14.40.Gx  (Properties of specific particles ) 
%           Mesons with S=C=B=0, mass > 2.5 GeV (including quarkonia)

%\keywords{}

\maketitle


%%%%%%%%%%%%%%%%%%%%%%%%%%%%%%%%%%%%%%%%%%%%%%%%%%%%%%%%%%%%%%%%%%%%%%%%%%%%%%
\section{Introduction\label{intro}}
%%%%%%%%%%%%%%%%%%%%%%%%%%%%%%%%%%%%%%%%%%%%%%%%%%%%%%%%%%%%%%%%%%%%%%%%%%%%%%

For a number of years, one of the largest discrepancies in the standard
model has been the disagreement between theory and experiment for
exclusive double-charmonium process $e^+e^-\to J/\psi+\eta_c$ at the 
$B$-factory energy of $10.58$~GeV. Initially,
the Belle Collaboration reported for the cross section times the
branching fraction into four or more charged tracks $\sigma[e^+e^-\to
J/\psi+\eta_c]\times B_{\ge 4}=33_{-6}^{+7}\pm 9~\hbox{fb}$
(Ref.~\cite{Abe:2002rb}). The first theoretical predictions were based
on NRQCD factorization calculations
\cite{Bodwin:1994jh} at leading order in $\alpha_s$, the QCD
coupling constant, and $v$, the heavy-quark (or antiquark) velocity in
the quarkonium rest frame. These predictions were $\sigma[e^+e^-\to
J/\psi+\eta_c]=3.78\pm 1.26~\hbox{fb}$ (Ref.~\cite{Braaten:2002fi}) and
$\sigma[e^+e^-\to J/\psi+\eta_c]=5.5~\hbox{fb}$
(Ref.~\cite{Liu:2002wq}).\footnote{The authors of
Ref.~\cite{Braaten:2002fi} initially reported a cross section of
$2.31\pm 1.09~\hbox{fb}$, but later corrected a sign error in the QED
interference term to arrive at the value cited above.} The calculation of
Ref.~\cite{Braaten:2002fi} includes QED effects, while that of
Ref.~\cite{Liu:2002wq} does not. Other differences between these
calculations arise from different choices of the charm-quark mass $m_c$,
NRQCD matrix elements, and $\alpha_s$. The sensitivities of the
calculations to these choices are indicative of large sources of
uncertainty in the theoretical calculations that have not yet been
quantified.

More recently, the Belle Collaboration has measured the production cross
section times the branching fraction into more than two charged
tracks and finds that $\sigma[e^+e^-\to J/\psi+\eta_c]\times
B_{>2}=25.6\pm 2.8\pm 3.4~\hbox{fb}$ (Ref.~\cite{Abe:2004ww}). The BABAR
Collaboration has also measured this quantity, and obtains
$\sigma[e^+e^-\to J/\psi+\eta_c]\times B_{> 2}=17.6\pm 2.8\pm
2.1~\hbox{fb}$ (Ref.~\cite{Aubert:2005tj}). These new experimental
results have narrowed the gap between theory and experiment.

An important recent theoretical development is the calculation of the
corrections of next-to-leading order (NLO) in $\alpha_s$
(Ref.~\cite{Zhang:2005ch}). These yield a $K$~factor of about $1.96$.
While this $K$~factor is substantial, it does not, by itself, completely
remove the discrepancy between theory and experiment.

Relativistic corrections to $\sigma[e^+e^-\to J/\psi+\eta_c]$ also 
make a significant contribution to the theoretical prediction. These
corrections  arise in two ways. First, they appear directly in the
corrections of order $v^2$ and higher to the process $e^+e^-\to J/\psi
+\eta_c$ itself. Second, they arise indirectly when one makes use of
phenomenological determinations of certain NRQCD matrix elements that 
appear in the expression for $\sigma[e^+e^-\to J/\psi +\eta_c]$. 
For example, the relevant matrix element of leading order in $v$
for the $J/\psi$ can be determined phenomenologically from the
experimental value for the width for $J/\psi \to e^+e^-$ and the
theoretical expression for that process. However, the theoretical
expression contains relativistic corrections, which then indirectly
affect the calculation of $\sigma[e^+e^-\to J/\psi +\eta_c]$.
The first relativistic correction appears at order $v^2$. ($v^2\approx
0.3$ for charmonium.) In Ref.~\cite{Braaten:2002fi}, the order-$v^2$
correction was calculated and was found to be about 
$1.95 \langle v^2\rangle_{J/\psi}+ 2.37 \langle v^2\rangle_{\eta_c}$.
Here, $\langle v^2\rangle_H$ is the ratio of an
order-$v^2$ nonperturbative NRQCD matrix element to the leading-order
matrix element in the quarkonium state
$H$. 
The large coefficients in the order-$v^2$
correction potentially lead to a relativistic correction. In
Ref.~\cite{Braaten:2002fi}, the $K$~factor for the relativistic
corrections was found to be $2.0_{-1.1}^{+10.9}$. The large
uncertainties arose from large uncertainties in the NRQCD matrix
elements.

Recently, progress has been made in reducing the uncertainties in the
order-$v^2$ NRQCD matrix elements by making use of a potential model to
calculate the quarkonium wave function \cite{Bodwin:2006dn}. The results
of Ref.~\cite{Bodwin:2006dn} allow one to make a meaningful
prediction for the relativistic corrections to $\sigma[e^+e^-\to
J/\psi+\eta_c]$. Making use of these results 
to compute the relativistic corrections and
taking into account the corrections of NLO in $\alpha_s$, the authors of
Ref.~\cite{Bodwin:2006ke} have given the prediction $\sigma[e^+e^-\to
J/\psi +\eta_c]=17.5\pm 5.7$~fb.

The authors of Ref.~\cite{He:2007te} have taken a different approach,
determining the NRQCD matrix elements of leading order in $v$ and of
relative order $v^2$ by using $\Gamma[J/\psi\to e^+e^-]$,             
$\Gamma[\eta_c\to\gamma\gamma]$, and $\Gamma[J/\psi\to \textrm{light
hadrons}]$ as inputs. Their result, $\sigma[e^+e^-\to J/\psi
+\eta_c]=20.04$~fb, is in agreement with the result of
Ref.~\cite{Bodwin:2006ke}. However, as we shall discuss, the values of
the individual matrix elements that were used in Ref.~\cite{He:2007te}
differ significantly from the values that were used in
Ref.~\cite{Bodwin:2006ke}.

The results of Refs.~\cite{Bodwin:2006ke} and \cite{He:2007te} suggest
that there is no longer a discrepancy between the experimental
measurements and the theoretical prediction. Nevertheless, it is useful
to include further refinements that improve the precision of the
theoretical prediction and to estimate as precisely as possible the
various theoretical uncertainties.

In the present paper, we carry out a new calculation of the relativistic
corrections to $\sigma[e^+e^-\to J/\psi +\eta_c]$. We include the
effects of pure QED processes, as well as QCD processes. In the case
of the pure QED processes, we incorporate a further refinement by
making use of the vector-meson-dominance (VMD) formalism to compute the
photon-fragmentation contribution. This approach reduces the theoretical
uncertainties that are associated with the pure QED contribution.
In our calculation, we make use of the approach of
Ref.~\cite{Bodwin:2006dn} to resum a class of relativistic corrections
to all orders in $v$. We also compute the contribution that arises from
the interference between the relativistic corrections and the
corrections of NLO in $\alpha_s$. Our calculation takes advantage of the
new higher-precision determinations of the relevant NRQCD matrix
elements in Ref.~\cite{BCKLY}. We make use of the detailed error
analysis of Ref.~\cite{BCKLY} to estimate the theoretical uncertainties
in our calculation, some of which are highly correlated.

The remainder of this paper is organized as follows. In
Sec.~\ref{sec:general-amp}, we discuss the general form of the amplitude
for the process $e^+e^-\to J/\psi +\eta_c$, and, in
Sec.~\ref{sec:gamma-Q4}, we discuss the corresponding quark-level
amplitude. Sec.~\ref{sec:nrqcd_exp} contains the expression for the
NRQCD expansion of the amplitude and a discussion of the matching
between NRQCD and full QCD. In Sec.~\ref{sec:resummation}, we describe
the resummation method that we employ. Sec.~\ref{sec:frame} contains
the specifics of the frame and coordinate system that we use in the
calculation. We present explicit formulas for the cross section in
Sec.~\ref{sec:x-section}. The VMD method for computing the fragmentation
contribution to the pure QED amplitude is summarized in
Sec.~\ref{sec:VMD}. We specify how we choose quarkonium masses in the
calculation in Sec.~\ref{sec:masses}. We present the method that we use
to compute the interference between the relativistic corrections and the
corrections of NLO in $\alpha_s$ in Sec.~\ref{sec:interference}. We give
our numerical results in Sec.~\ref{sec:results} and compare them with
the results from previous calculations in Sec.~\ref{sec:comparison}.
Finally, we summarize and discuss our results in
Sec.~\ref{sec:discussion}.

%%%%%%%%%%%%%%%%%%%%%%%%%%%%%%%%%%%%%%%%%%%%%%%%%%%%%%%%%%%%%%%%%%%%%%%%%%%%%%
\section{General Form of the amplitude for 
$\bm{e^+e^-\to J/\psi+\eta_c}$
\label{sec:general-amp}}
%%%%%%%%%%%%%%%%%%%%%%%%%%%%%%%%%%%%%%%%%%%%%%%%%%%%%%%%%%%%%%%%%%%%%%%%%%%%%%

Let us consider the amplitude for the exclusive process
$\gamma^\ast\to J/\psi(P_1,\lambda)+\eta_c(P_2)$, where $\lambda$ is the
helicity of the $J/\psi$. The $S$-matrix element for 
$e^+(k_2) e^-(k_1)\to J/\psi(P_1,\lambda)+\eta_c(P_2)$
can be written as
%------------------
\begin{equation}
%------------------
\mathcal{M}(\lambda)=L^\mu \,\mathcal{A}_\mu[J/\psi(\lambda)+\eta_c],
\label{em}%
%------------------
\end{equation}
%------------------
where the leptonic factor $L^\mu$ is defined by
%------------------
\begin{equation}
%------------------
L^\mu=-i\frac{e_c e^2}{s}\bar{v}(k_2)\gamma^\mu u(k_1).
\label{L-mu}%
%------------------
\end{equation}
%------------------
Here, $e_c$ is the electric charge of the charm quark and 
$s=4E_{\textrm{beam}}^2$ is the square of the $e^+e^-$ 
center-of-momentum (CM) energy. 
$\mathcal{A}_\mu[J/\psi(\lambda)+\eta_c]$ in Eq.~(\ref{em}) is
the vacuum-to-$J/\psi+\eta_c$ matrix element. It can be expressed in 
the following form, which derives from the Lorentz invariance of the 
amplitude and the parity conservation of the strong and electromagnetic 
interactions~\cite{Braaten:2002fi}:
%------------------
\begin{equation}
%------------------
\mathcal{A}_\mu[J/\psi(\lambda)+\eta_c]
=
\langle J/\psi(P_1,\lambda)+\eta_c(P_2)|J_\mu(0)|0\rangle
=
iA\epsilon_{\mu\nu\alpha\beta}P_1^\nu P_2^\alpha 
\epsilon^{\ast\beta}(\lambda),
%------------------
\label{me}%
\end{equation}
%------------------
where $J_\mu(0)$ is the electromagnetic current, 
$\epsilon^\ast(\lambda)$ is the polarization four-vector of the
$J/\psi$ with helicity $\lambda$ whose components in the $J/\psi$ 
rest frame are $\epsilon^\ast(\lambda)=[0,\bm{\epsilon}^\ast(\lambda)]$.
The convention 
for the antisymmetric tensor in Eq.~(\ref{me}) is chosen so that 
$\epsilon_{0123}=+1$. We note that $A$ is parity invariant and that 
$\mathcal{A}^\mu$ transforms as a four-vector under parity. 

%%%%%%%%%%%%%%%%%%%%%%%%%%%%%%%%%%%%%%%%%%%%%%%%%%%%%%%%%%%%%%%%%%%%%%%%%%%%%%
\section{Amplitude for $\bm{\gamma^\ast\to} \bm{Q}\bar{\bm{Q}}
\bm{({}^3S_1)} \bm{+} \bm{Q}\bar{\bm{Q}}\bm{({}^1S_0)}$
\label{sec:gamma-Q4}}
\begin{figure}
\centerline{
\includegraphics[height=3.5cm]{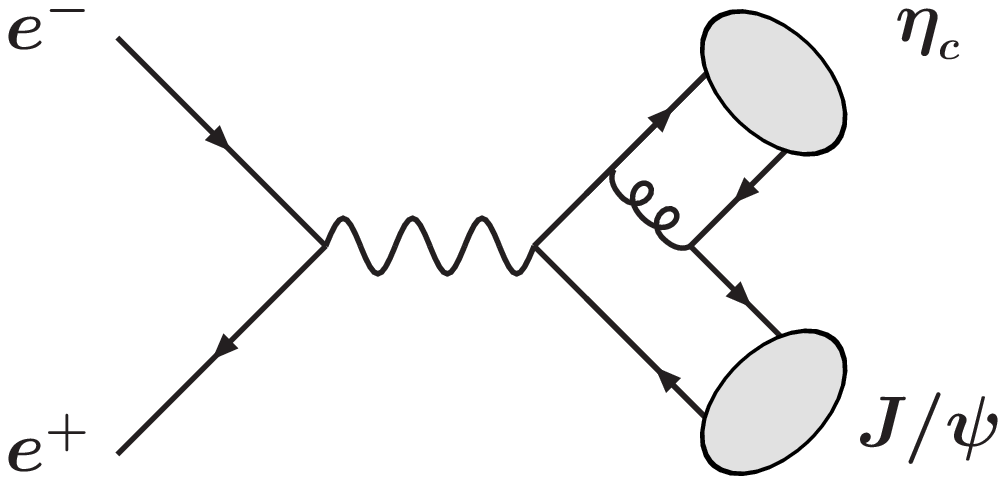}
\quad\quad
\includegraphics[height=3.5cm]{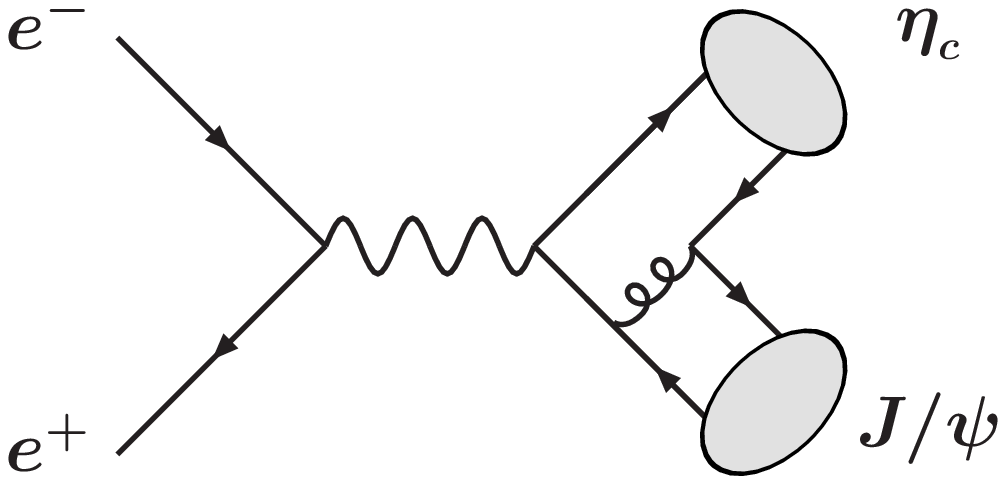}
}
\caption{Feynman diagrams for the process $e^+ e^- \to J/\psi + \eta_c$
at leading order in $\alpha_s$. The wavy line represents a photon, the 
curly line represents a gluon, and the straight lines represent 
the leptons and heavy quarks. There are six additional diagrams that
can be obtained by reversing the directions of the arrows on the
heavy-quark lines and/or by replacing the gluon by a photon.}
\label{fig1}
\end{figure}

\begin{figure}
\centerline{  
\includegraphics[height=3.5cm]{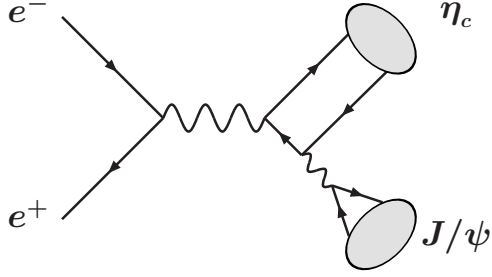}
}                                                 
\caption{Feynman diagram for the process $e^+ e^- \to J/\psi + \eta_c$
in which a photon fragments into a $J/\psi$. 
There is an additional diagram that can be obtained by reversing the 
directions of the arrows on the heavy-quark line on the $\eta_c$ side.}  
\label{fig2}     
\end{figure}

%%%%%%%%%%%%%%%%%%%%%%%%%%%%%%%%%%%%%%%%%%%%%%%%%%%%%%%%%%%%%%%%%%%%%%%%%%%%%%
The exclusive process $\gamma^\ast\to J/\psi(P_1,\lambda)+\eta_c(P_2)$
involves the decay of a virtual photon into two heavy quark-antiquark
($Q\bar Q$) pairs $Q(p_i)\bar{Q}(\bar{p}_i)$ ($i=1\hbox{~or~}2$). Both
of the pairs are in color-singlet states. At leading order in
$\alpha_s$, the process proceeds through the diagrams shown in
Fig.~\ref{fig1}, plus two additional diagrams in which the directions of
the arrows on the heavy-quark lines are reversed.
There are also purely electromagnetic contributions to the process. At
leading order in the QED coupling $\alpha$, two types of diagrams
contribute to the QED processes. The first type consists of the diagrams
shown in Fig.~\ref{fig1} (plus two others in which the directions of
the arrows on the heavy-quark lines are reversed), but with the gluon
replaced by a photon. The second type consists of diagrams in which a
photon fragments into a $J/\psi$. One of these diagrams is shown in
Fig.~\ref{fig2}. (There is an additional diagram in which the 
directions of the arrows on the heavy-quark line on the $\eta_c$ side 
are reversed.)

%%%%%%%%%%%%%%%%%%%%%%%%%%%%%%%%%%%%%%%%%%%%%%%%%%
\subsection{Kinematics\label{sec:kinematics}}
%%%%%%%%%%%%%%%%%%%%%%%%%%%%%%%%%%%%%%%%%%%%%%%%%%

The $Q$ and $\bar{Q}$ that evolve into the charmonium $H_i$ with
momentum $P_i$ have momenta $p_i$ and $\bar{p}_i$, where  $i=1$
denotes the $J/\psi$ and $i=2$ denotes the  $\eta_c$.
The $Q(p_1)\bar{Q}(\bar{p}_1)$ pair is in a spin-triplet $S$-wave state and
the $Q(p_2)\bar{Q}(\bar{p}_2)$ pair is in a spin-singlet $S$-wave state. 
The four-momenta of the $Q$ and $\bar Q$ in the $i$-th pair 
are expressed in terms of the total momentum $P_i$ and the relative momentum 
$q_i$:
%------------------
\begin{subequations}
\begin{eqnarray}
%------------------
p_i&=&\tfrac{1}{2}P_i+q_i,
\\
\bar{p}_i&=&\tfrac{1}{2}P_i-q_i.
%------------------
\end{eqnarray}
\label{p-i}%
\end{subequations}
%------------------
$P_i$ and $q_i$ are chosen to be orthogonal: $P_i\cdot q_i=0$.
In the rest frame of the $i$-th $Q\bar{Q}$ pair, the explicit components 
of the momenta listed above are $P_i=[2E(q_i),\bm{0}]$, $q_i=(0,\bm{q}_i)$,
$p_i=[E(q_i),\bm{q}_i]$, 
and $\bar{p}_i=[E(q_i),-\bm{q}_i]$, 
respectively, where $E(q_i)=\sqrt{m_c^2+\bm{q}^2_i}$ is the energy of 
the $Q$ or the $\bar{Q}$ in the $Q\bar{Q}$ rest frame.

%%%%%%%%%%%%%%%%%%%%%%%%%%%%%%%%%%%%%%%%%%%%%%%%%%
\subsection{Spin and color projectors\label{sec:spin-projection}}
%%%%%%%%%%%%%%%%%%%%%%%%%%%%%%%%%%%%%%%%%%%%%%%%%%

A production amplitude of a $Q(p_i)\bar{Q}(\bar{p_i})$ pair 
can be expressed in the form
%------------------
\begin{equation}
%------------------
\bar{u}(p_i) \mathcal{A}  v(\bar{p}_i) =
\textrm{Tr} \big[ \mathcal{A} \, v(\bar{p}_i) \bar{u}(p_i) \big],
%------------------
\label{vbarAu}%
\end{equation}
%------------------
where $\mathcal{A}$ is a matrix that acts on spinors with both
Dirac and color indices.  The amplitude in Eq.~(\ref{vbarAu})
can be projected into a particular spin and color channel by 
replacing $v(\bar{p}_i)\bar{u}(p_i)$ with a projection matrix. 
The color projector $\pi_1$ onto a color-singlet state is
%------------------
\begin{equation}
%------------------
\pi_1=\frac{1}{\sqrt{N_c}} \mathbbm{1},
\label{color-projector}%
\end{equation}
%------------------
where $\mathbbm{1}$ is the $3\times 3$ unit matrix of the fundamental 
representation of SU(3). The color-singlet projector 
(\ref{color-projector}) is normalized so that 
Tr[$\pi_1 \pi_1^\dagger$]=1. The projector of the pair 
$Q(p_1)\bar{Q}(\bar{p}_1)$ onto a spin-triplet state with helicity $\lambda$ 
and the 
projector of the pair $Q(p_2)\bar{Q}(\bar{p}_2)$ onto a spin-singlet 
state 
are denoted by $\Pi_3(p_1,\bar{p}_1,\lambda)$ and 
$\Pi_1(p_2,\bar{p}_2)$, respectively. The projectors, valid to all orders 
in $\bm{q}_i$, are given in Ref.~\cite{Bodwin:2002hg}:
%------------------
\begin{subequations}
\begin{eqnarray}
%------------------
\Pi_3(p_1,\bar{p}_1,\lambda)&=&
-\frac{1}{4\sqrt{2}E(q_1)[E(q_1)+m_c\,]}
(/\!\!\!\bar{p}_1-m_c)
\,/\!\!\!\epsilon^{\,\ast}(\lambda)
[\,/\!\!\!\!P_1\!+\!2E(q_1)]
(/\!\!\!{p}_1+m_c),
\\
\Pi_1(p_2,\bar{p}_2)&=&
\frac{1}{4\sqrt{2}E(q_2)[E(q_2)+m_c\,]}
(/\!\!\!\bar{p}_2-m_c)
\gamma^5
[\,/\!\!\!\!P_2\!+\!2E(q_2)]
(/\!\!\!{p}_2+m_c),
%------------------
\end{eqnarray}
\label{spin-projector}%
\end{subequations}
%------------------
where the spin-polarization vector $\epsilon^\ast(\lambda)$ 
satisfies $P_1 \cdot \epsilon^\ast(\lambda)=0$.
The spin projectors in Eq.~(\ref{spin-projector}) are normalized so that
%------------------
\begin{subequations}
\begin{eqnarray}
%------------------
\textrm{Tr}[\Pi_3(p_1,\bar{p}_1,\lambda)
            \Pi_3^\dagger(p_1,\bar{p}_1,\lambda)] 
&=& 4 p^0_1 \bar{p}^0_1,
\\
\textrm{Tr}[\Pi_1(p_2,\bar{p}_2)  \Pi^\dagger_1(p_2,\bar{p}_2)]
&=& 4 p^0_2 \bar{p}^0_2.
%------------------
\end{eqnarray}
\label{norm-projector}%
\end{subequations}
%------------------
Since we are considering an exclusive process, in which no hadrons are
present other than the $J/\psi$ and the $\eta_c$, we consider only the
states of the $Q\bar{Q}$ pairs that have the same quantum numbers as the
$J/\psi$ and the $\eta_c$. That is, the pair
$Q(p_1)\bar{Q}(\bar{p}_1)$ must be in a color-singlet, spin-triplet
$S$-wave state, as is the  $J/\psi$, and the pair
$Q(p_2)\bar{Q}(\bar{p}_2)$ must be in a color-singlet, spin-singlet
$S$-wave state, as is the $\eta_c$.

The spin projectors in Eq.~(\ref{spin-projector}) can be 
simplified as follows:
%--------------
\begin{subequations}
\begin{eqnarray}
\Pi_3(p_1,\bar{p}_1,\lambda)&=&
-\frac{1}{2\sqrt{2}E(q_1)}
\left(\,/\! \! \! \bar{p}_1- m_c  \right)
\left(
\,/\!\!\!\epsilon^{\,\ast}(\lambda)
-\frac{(p_1-\bar{p}_1\,)\cdot \epsilon^\ast(\lambda)}
{2[E(q_1)+m_c\,]}
\right)
\left(\,/\! \! \!           p_1 + m_c  \right),
\\
\Pi_1(p_2,\bar{p}_2)&=&
\frac{1}{2\sqrt{2}E(q_2)}
\left(\,/\! \! \! \bar{p}_2- m_c  \right)
\gamma_5
\left(\,/\! \! \!           p_2 + m_c  \right).
\end{eqnarray}
\label{simple}%
\end{subequations}
%--------------
The spin
projectors in Eq.~(\ref{simple}) are also valid to all orders in
$\bm{q}_i$.

In addition, we provide the following formulas, valid to all order in 
$\bm{q}_i$, which are useful in this calculation:
%--------------
\begin{subequations}
\begin{eqnarray}
\gamma_\alpha\Pi_3(p_1,\bar{p}_1,\lambda)\gamma^\alpha
&=&
\frac{1}{\sqrt{2}E(q_1)}
\Bigg[
/\! \! \! {p}_1
\,/\!\!\!\epsilon^{\,\ast}(\lambda) 
/\! \! \! \bar{p}_1
-m_c^2\,/\!\!\!\epsilon^{\,\ast}(\lambda)
+
(p_1-\bar{p}_1\,)\cdot \epsilon^\ast(\lambda)
\nonumber\\
&&
\quad \quad \quad
\quad \quad 
\times
\left(
2E(q_1)+\frac{m_c(/\! \! \! p_1 - /\! \! \! \bar{p}_1  \,)}
{2[E(q_1)+m_c\,]}
\right)
\Bigg],
\\
\gamma_\alpha\Pi_1(p_2,\bar{p}_2)\gamma^\alpha&=&
\frac{1}{\sqrt{2}}\;\gamma_5
\left[4E(q_2)-\frac{m_c}{E(q_2)}\,/\!\!\!\!P_2\right].
\end{eqnarray}
\label{simple-sum}%
\end{subequations}
%--------------

%%%%%%%%%%%%%%%%%%%%%%%%%%%%%%%%%%%%%%%%%%%%%%%%%%
\subsection{Projections of the four-quark
states\label{sec:s-wave-projection}}
%%%%%%%%%%%%%%%%%%%%%%%%%%%%%%%%%%%%%%%%%%%%%%%%%%

From the full QCD amplitude for 
$\gamma^\ast\to Q(p_1)\bar{Q}(\bar{p}_1)Q(p_2)\bar{Q}(\bar{p}_2)$,
one can project out the amplitude for 
$\gamma^\ast\to Q\bar{Q}(P_1,q_1,\lambda)+Q\bar{Q}(P_2,q_2)$,
where the first pair is in a color-singlet, spin-triplet state
with helicity $\lambda$ and the second pair is in a color-singlet,
spin-singlet state. Applying the spin projections
to both $Q\bar{Q}$ pairs simultaneously, one obtains
%------------------
\begin{equation}
%------------------
\mathcal{A}_Q^{\mu}(P_1,q_1,\lambda;P_2,q_2)
=
\textrm{Tr}\big\{\mathcal{A}^\mu
[\gamma^\ast\to Q(p_1)\bar{Q}(\bar{p}_1)Q(p_2)\bar{Q}(\bar{p}_2)]
\,
  [\Pi_3(p_1,\bar{p}_1,\lambda) \otimes \pi_1]
\otimes
  [\Pi_1(p_2,\bar{p}_2) \otimes \pi_1] \big\},
%------------------
\label{amp-spin-double}%
\end{equation}
%------------------
where 
$\mathcal{A}^\mu[\gamma^\ast\to Q(p_1)\bar{Q}(\bar{p}_1)%
Q(p_2)\bar{Q}(\bar{p}_2)]$ is the full QCD amplitude for 
$\gamma^\ast\to Q(p_1)\bar{Q}(\bar{p}_1)Q(p_2)\bar{Q}(\bar{p}_2)$,
$\mu$ is the vector index of the virtual photon, and
$\mathcal{A}_Q^{\mu}(P_1,q_1,\lambda;P_2,q_2)$ is the amplitude
for $\gamma^\ast\to Q\bar{Q}(P_1,q_1,\lambda)+Q\bar{Q}(P_2,q_2)$.

In the amplitude (\ref{amp-spin-double}), the $Q\bar Q$ pairs are not
necessarily in $S$-wave orbital-angular-momentum states. One can
project out the $S$-wave amplitude by averaging, for each $Q\bar Q$
pair, over the direction of the relative momentum $q_i$ in the
$Q\bar{Q}(P_i)$ rest frame. The amplitude for $\gamma^\ast\to
Q\bar{Q}({}^3S_1,P_1,\lambda) + Q\bar{Q}({}^1S_0,P_2)$ is
%------------------
\begin{equation}
%------------------
\mathcal{A}_Q^{\mu}({}^3S_1,P_1,\lambda;{}^1S_0,P_2)
=
\overline{\mathcal{A}_Q^{\mu}(P_1,q_1,\lambda;P_2,q_2)},
%------------------
\label{amp-QQ-fin}%
\end{equation}
%------------------
where the bar on the right side of  
Eq.~(\ref{amp-QQ-fin}) is the average over the angles of both $q_1$
and $q_2$ in the $P_1$ and $P_2$ rest frames, respectively:
%------------------
\begin{equation}
%------------------
\overline{\mathcal{A}_Q^{\mu}(P_1,q_1,\lambda;P_2,q_2)}
=\int
\frac{d\Omega_1 d\Omega_2}{(4\pi)^2}
\mathcal{A}_Q^{\mu}(P_1,q_1,\lambda;P_2,q_2).
%------------------
\label{amp-QQ-angle}%
\end{equation}
%------------------
$d\Omega_i$ is the solid-angle element of $q_i$, defined in
the $P_i$ rest frame. Once we have averaged over angles, the $\bm{q}_i$ 
dependence in the amplitude (\ref{amp-QQ-fin}) reduces a dependence only 
on $\bm{q}_1^2$ and $\bm{q}^2_2$.
Note that $P_i$ depends on $\bm{q}_i^2$ implicitly: 
$P_i^2=4(m_c^2+\bm{q}_i^2)$.

%%%%%%%%%%%%%%%%%%%%%%%%%%%%%%%%%%%%%%%%%%%%%%%%%%%%%%%%%%%%%%%%%%%%%%%%%%%%%%
\section{NRQCD expansion of the amplitude and matching
\label{sec:nrqcd_exp}}
%%%%%%%%%%%%%%%%%%%%%%%%%%%%%%%%%%%%%%%%%%%%%%%%%%%%%%%%%%%%%%%%%%%%%%%%%%%%%%

The NRQCD expansion of  Eq.~(\ref{me}) in terms of  the vacuum-to-$J/\psi$ 
and vacuum-to-$\eta_c$ matrix elements  is
\begin{equation}
\mathcal{A}^\mu[J/\psi(\lambda)+\eta_c]=
\sqrt{2m_1}
\sqrt{2m_2}
\sum_{m,n}
d_{mn}^\mu\langle J/\psi(\lambda)|\mathcal{O}_m| 0\rangle
\langle \eta_c|\mathcal{O}_n|0\rangle,
\label{nrqcd-exp}
\end{equation}
where the $d_{mn}^\mu$ are short-distance coefficients and the
$\mathcal{O}_m$ and the $\mathcal{O}_n$ are NRQCD operators. The quantities
$m_1$ and $m_2$ represent the $J/\psi$ and $\eta_c$ masses,
respectively. However, as we shall discuss later, these quantities are
not necessarily equal to the physical meson masses, but may
instead be expressed as functions of the heavy-quark masses via the
nonrelativistic expansion of NRQCD. The factor $\sqrt{2m_1}\sqrt{2m_2}$
appears on the right side of Eq.~(\ref{nrqcd-exp}) because we use
relativistic normalization for the meson states in
$\mathcal{A}^\mu[J/\psi(\lambda)+\eta_c]$, but we use conventional
nonrelativistic normalization for the NRQCD matrix elements on the right
side of Eq.~(\ref{nrqcd-exp}).

Now we approximate the formula (\ref{nrqcd-exp}) by retaining only those 
operator matrix elements that connect the vacuum to the 
color-singlet $Q\bar Q$ Fock states of the quarkonia. Then, we 
have
%------------------
\begin{eqnarray}
%------------------
\mathcal{A}^\mu[J/\psi(\lambda)+\eta_c]
&=&
\sqrt{2m_1}
\sqrt{2m_2}
\sum_{m=0}^\infty
\sum_{n=0}^\infty
c^\mu_{mn}(\lambda)
\langle J/\psi(\lambda)|
\psi^\dagger (-\tfrac{i}{2}\tensor{\bm{D}})^{2m} 
\bm{\sigma}\cdot\bm{\epsilon}(\lambda)\chi|0\rangle
\nonumber\\
&&
\quad\quad\quad
\quad\quad\quad
\quad\quad\quad
\times
\langle \eta_c|\psi^\dagger 
(-\tfrac{i}{2}\tensor{\bm{D}})^{2n}\chi|0\rangle,
%------------------
\label{A-HH}%
\end{eqnarray}
%------------------
where the short-distance coefficients $c^\mu_{mn}(\lambda)$ are a subset
of the short-distance coefficients $d_{mn}^\mu$. $\psi^\dagger$ and
$\chi$ are two-component Pauli spinors that create a heavy quark and a
heavy antiquark, respectively, $\sigma^i$ is a Pauli matrix, and
$\tensor{\bm{D}}$ is the spatial part of the covariant derivative acting
to the left and right anti-symmetrically. Note that there is no sum
over $\lambda$ on the right side of Eq.~(\ref{A-HH}). All of the
three-vector quantities in the NRQCD matrix elements for the $H_i$ are
defined in the $P_i$ rest frame. We will clarify below the meaning of
the approximation that we have taken to arrive at Eq.~(\ref{A-HH}).

The short-distance coefficients $c^\mu_{mn}(\lambda)$ can be 
obtained from the full QCD amplitude 
$\mathcal{A}^\mu_{Q}({}^3S_1,P_1,\lambda;{}^1S_0,P_2)$ in 
Eq.~(\ref{amp-QQ-fin}).
The NRQCD expansion of the full QCD amplitude is 
%------------------
\begin{eqnarray}
%------------------
\mathcal{A}^\mu_{Q}({}^3S_1,P_1,\lambda;{}^1S_0,P_2)
&=&
\sum_{m=0}^\infty
\sum_{n=0}^\infty
\,\,
c_{mn}^\mu(\lambda)
\langle Q\bar{Q}({}^3S_1,\lambda)|
\psi^\dagger (-\tfrac{i}{2}\tensor{\bm{D}})^{2m}
\bm{\sigma}\cdot\bm{\epsilon}(\lambda)
\chi|0\rangle
\nonumber\\
&&
\quad \quad \quad
\quad \quad \quad
\times
\langle Q\bar{Q}({}^1S_0)|\psi^\dagger 
(-\tfrac{i}{2}\tensor{\bm{D}})^{2n}\chi|0\rangle
\nonumber\\
&=&
8N_cE(q_1)E(q_2)
\sum_{m=0}^\infty
\sum_{n=0}^\infty
c^\mu_{mn}(\lambda) \bm{q}_1^{2m}\bm{q}_2^{2n}.
\label{A-QQ}%
%------------------
\end{eqnarray}
%------------------
In Eq.~(\ref{A-QQ}), we use relativistic normalization for the $Q$ and
$\bar Q$ states in the computation of
$\mathcal{A}^\mu_Q({}^3S_1,P_1,\lambda;{}^1S_0,P_2)$ and in the
computation of the NRQCD matrix elements. Consequently, a factor
$4E(q_1)E(q_2)$ appears in the second equality of Eq.~(\ref{A-QQ}). An
additional factor $2N_c$ arises from the spin and color factors of the
NRQCD matrix elements. From Eq.~(\ref{A-QQ}), it is straightforward to
calculate the short-distance coefficients $c_{mn}^\mu(\lambda)$:
%------------------
\begin{equation}
%------------------
c^\mu_{mn}(\lambda)=\frac{1}{m!\, n!}
\frac{\partial^m}{\partial \bm{q}_1^{2m}}
\frac{\partial^n}{\partial \bm{q}_2^{2n}}
\left.
\left[
\frac{\mathcal{A}^\mu_{Q}({}^3S_1,P_1,\lambda;{}^1S_0,P_2)}
{8N_cE(q_1)E(q_2)}
\right] \right|_{\bm{q}_1^2=\bm{q}_2^2=0}. 
\label{cmn-fin}
%------------------
\end{equation}
%------------------

Substituting the short-distance coefficients (\ref{cmn-fin}) 
into Eq.~(\ref{A-HH}), one finds that 
%------------------
\begin{eqnarray}
%------------------
\mathcal{A}^\mu[J/\psi(\lambda)+\eta_c]
&=&
\sqrt{2m_1}
\sqrt{2m_2}
\langle J/\psi(\lambda)|\psi^\dagger 
\bm{\sigma}\cdot\bm{\epsilon}(\lambda)\chi|0\rangle
\langle \eta_c|\psi^\dagger\chi|0\rangle
\nonumber\\&&
\times
\sum_{m=0}^\infty
\sum_{n=0}^\infty
c^\mu_{mn}(\lambda)
\langle \bm{q}^{2m}\rangle_{J/\psi}
\langle \bm{q}^{2n}\rangle_{\eta_c}
\nonumber\\
&=&
\frac{\sqrt{2m_1}\sqrt{2m_2}}{2N_c}
\langle J/\psi(\lambda)|\psi^\dagger 
\bm{\sigma}\cdot\bm{\epsilon}(\lambda)\chi|0\rangle
\langle \eta_c|\psi^\dagger\chi|0\rangle
\sum_{m=0}^\infty
\sum_{n=0}^\infty
\frac{\langle \bm{q}^{2m}\rangle_{J/\psi}
\langle \bm{q}^{2n}\rangle_{\eta_c}}{m!\,n!}
\nonumber\\
&&
\times
\left.
\left(\frac{\partial}{\partial \bm{q}_1^{2}}\right)^m
\left(\frac{\partial}{\partial \bm{q}_2^{2}}\right)^n
\left[
\frac{
\mathcal{A}_{Q}^\mu({}^3S_1,P_1,\lambda;{}^1S_0,P_2)} {4E(q_1)E(q_2)}
\right] \right|_{\bm{q}_1^2=\bm{q}_2^2=0}.
%------------------
\label{A-HH2}%
\end{eqnarray}
%------------------
Here, the quantities $\langle \bm{q}^{2m}\rangle_H$ are ratios of
NRQCD matrix elements:
%------------------
\begin{subequations}
\begin{eqnarray}
\langle \bm{q}^{2m}\rangle_{J/\psi}&=&
\frac{\langle J/\psi(\lambda)|
\psi^\dagger (-\tfrac{i}{2}\tensor{\bm{D}})^{2m} 
\bm{\sigma}\cdot\bm{\epsilon}(\lambda)\chi|0\rangle}
{\langle J/\psi(\lambda)|
\psi^\dagger \bm{\sigma}\cdot\bm{\epsilon}(\lambda)\chi|0\rangle},
\label{q2m-psi}
\\
\langle \bm{q}^{2n}\rangle_{\eta_c}&=&
\frac{\langle \eta_c|\psi^\dagger 
(-\tfrac{i}{2}\tensor{\bm{D}})^{2n} \chi|0\rangle}
{\langle \eta_c|\psi^\dagger \chi|0\rangle}.
\end{eqnarray}
\end{subequations}
%------------------
We note that $\langle J/\psi(\lambda)|\psi^\dagger 
\bm{\sigma}\cdot\bm{\epsilon}(\lambda)\chi|0\rangle$ and $\langle 
\bm{q}^{2m}\rangle_{J/\psi}$ are independent of the $J/\psi$ helicity 
$\lambda$, and there are no sums over $\lambda$ in 
these quantities.

Now we can clarify the meaning of the approximation that was taken to
arrive at Eq.~(\ref{A-HH}) and, consequently, to arrive at
Eq.~(\ref{A-HH2}). Suppose that we specialize to the Coulomb gauge.
Then, we can drop the gauge fields in covariant derivatives in the
matrix elements in Eq.~(\ref{A-HH2}), making errors of relative order
$v^2$. The matrix elements are then proportional to derivatives of the
Coulomb-gauge color-singlet $Q\bar Q$ quarkonium wave function at the
origin \cite{Bodwin:1994jh}. That is, $\langle \bm{q}^{2n} \rangle$ is just
the $2n$th moment of the momentum-space wave function with respect to
the wave-function momentum (the relative momentum of the $Q$ and $\bar
Q$). Hence, Eq.~(\ref{A-HH2}) has the interpretation of the convolution
of the short-distance amplitude with the momentum-space quarkonium wave
functions, where the short-distance coefficients have been Taylor
expanded with respect to the wave-function momenta. Therefore, we see
that the approximate NRQCD expansion in Eqs.~(\ref{A-HH}) and
(\ref{A-HH2}) includes all of the relativistic corrections that are
contained in the color-singlet $Q\bar Q$ quarkonium wave function, up to
the ultraviolet cutoff of the NRQCD matrix elements.\footnote{We 
note that, in the case of dimensionally regulated NRQCD matrix elements, 
pure power ultraviolet divergences in the matrix elements are set to zero. 
Hence, the effects of pure-power-divergent contributions are absent in the 
resummation.}

%%%%%%%%%%%%%%%%%%%%%%%%%%%%%%%%%%%%%%%%%%%%%%%%%%%%%%%%%%%%%%%%%%%%%%%%%%%%%%
\section{Resummation \label{sec:resummation}}
%%%%%%%%%%%%%%%%%%%%%%%%%%%%%%%%%%%%%%%%%%%%%%%%%%%%%%%%%%%%%%%%%%%%%%%%%%%%%%

In Ref.~\cite{Bodwin:2006dn}, a method was presented for resumming a
class of relativistic corrections to the color-singlet
$S$-wave amplitudes that appear in the production and decay
of $S$-wave quarkonium states. The key to
the resummation is an expression that relates the $S$-wave color-singlet
matrix elements of higher orders in $v$ to the matrix elements of
relative orders $v^0$ and $v^2$:
%-------------------
\begin{equation}
%-------------------
\langle \bm{q}^{2n}\rangle_H
=\langle \bm{q}^{2}\rangle_H^n.
\label{master}%
%-------------------
\end{equation}
%-------------------
The relation (\ref{master}) is derived in the approximation in which the
$Q$ and $\bar Q$ interact only through the leading
spin-independent potential. Consequently, the relation
(\ref{master}) is accurate up to corrections of relative order
$v^2$.\footnote{The derivation involves specializing to the Coulomb 
gauge and replacing covariant derivatives in operators with ordinary 
derivatives. This approximation also introduces errors of relative order 
$v^2$.} 

The amplitude (\ref{A-HH2}) is a function of the ratios $\langle
\bm{q}^{2m}\rangle_{J/\psi}$ and $\langle \bm{q}^{2n}\rangle_{\eta_c}$.
Applying the relation (\ref{master}) to
Eq.~(\ref{A-HH}), one obtains the resummed expression 
%------------------
\begin{eqnarray}
%------------------
\mathcal{A}^\mu[J/\psi(\lambda)+\eta_c]
&=&
\frac{1}{2N_c}
\langle J/\psi(\lambda)|
\psi^\dagger \bm{\sigma}\cdot\bm{\epsilon}(\lambda)\chi|0\rangle
\langle \eta_c|\psi^\dagger \chi|0\rangle
\nonumber\\
&&\times
\left.\left[\frac{\sqrt{2m_1}\sqrt{2m_2}} {2E(q_1)\,2E(q_2)}
\mathcal{A}_{Q}^\mu({}^3S_1,P_1,\lambda;{}^1S_0,P_2)
\right]\right|_{
\bm{q}_1^2=\langle \bm{q}^2\rangle_{J/\psi},\,
\bm{q}_2^2=\langle \bm{q}^2\rangle_{\eta_c}}.
%------------------
\label{A-HH3}%
\end{eqnarray}
%------------------
The expression (\ref{A-HH3}) resums those relativistic corrections
that are contained in the $Q\bar Q$ quarkonium wave function in the
leading-potential model for the wave function. We note that, because
the relation (\ref{master}) is accurate only up to corrections of
relative order $v^2$, the use of the resummed expression (\ref{A-HH3})
generally does not improve the nominal accuracy over that which one
would obtain by retaining only corrections through relative order $v^2$
in Eq.~(\ref{A-HH}). The exception to this is the situation in which the
short-distance coefficients $c_{mn}^\mu$ in Eq.~(\ref{cmn-fin}) grow
rapidly with $m$ or $n$. Then the terms of nominally higher order in $v$
in Eq.~(\ref{A-HH}) can have numerical values that are comparable to or
larger than the numerical value of the term of nominal order $v^2$. In
that situation, the resummed expression can give an improved estimate of
the amplitude. The resummed expression may also give an indication of
the rate of convergence of the $v$ expansion. In any case, it is
generally useful to include a well-defined set of higher-order
contributions in a calculation whenever possible.

%%%%%%%%%%%%%%%%%%%%%%%%%%%%%%%%%%%%%%%%%%%%%%%%%%%%%%%%%%%%%%%%%%%%%%%%%%%%%%
\section{Choice of frame and co-ordinate system
\label{sec:frame}}
%%%%%%%%%%%%%%%%%%%%%%%%%%%%%%%%%%%%%%%%%%%%%%%%%%%%%%%%%%%%%%%%%%%%%%%%%%%%%%

In calculating $\mathcal{A}_{Q}^\mu({}^3S_1,P_1,\lambda;{}^1S_0,P_2)$,
it is convenient to specialize to the $e^+e^-$ CM frame, to choose a
particular coordinate system, and to choose a particular convention for
the polarization vectors of the $Q\bar{Q}_1({}^3S_1)$ states 
for the various helicities. We make these choices as follows:
%------------------
\begin{subequations}
\begin{eqnarray}
%------------------
k_1&=&\frac{\sqrt{s}}{2}
(1,+\sin\theta,0,+\cos\theta),
\\
k_2&=&\frac{\sqrt{s}}{2}
(1,-\sin\theta,0,-\cos\theta),
\\
P_1^\ast&=&(E_1,0,0,+P_{\textrm{CM}}),
\\
P_2^\ast&=&(E_2,0,0,-P_{\textrm{CM}}),
\\
\epsilon^{\ast}(0)&=&
\frac{1}{\sqrt{E_1^2-P_{\textrm{CM}}^2}}(P_{\textrm{CM}},0,0,E_1),
\\
\epsilon^{\ast}(\pm)&=&\mp\frac{1}{\sqrt{2}}(0,1,\mp i,0).
%------------------
\end{eqnarray}
\label{cmframe}%
\end{subequations}
%------------------
Here, the angle $\theta$ is the scattering angle, four-vectors 
are written as $v=(v^0,v^1,v^2,v^3)$, and
%------------------
\begin{subequations}
\begin{eqnarray}
%------------------
P_{\textrm{CM}}&=&
\frac{\lambda^{1/2}\big(s,\tilde{m}_1^2,\tilde{m}_2^2\big)}{2\sqrt{s}},
\\
E_i&=&
\sqrt{
P_{\textrm{CM}}^2+\tilde{m}_i^2},
%------------------
\end{eqnarray}
\label{PCM}%
\end{subequations}
%------------------
where
$\lambda(x,y,z)=x^2+y^2+z^2-2(xy+yz+zx)$.
We have used the notation $P_i^\ast$ to distinguish the particular 
values of these quantities in the $e^+e^-$ CM frame from the values in 
the quarkonium rest frame (Sec.~\ref{sec:kinematics}).
As with the quantities $m_1$ and $m_2$ in Eq.~(\ref{nrqcd-exp}), 
$\tilde{m}_1$ and $\tilde{m}_2$ represent the $J/\psi$ 
and $\eta_c$ masses, respectively. We will specify below how these are 
chosen for various parts of the calculation. 

Now let us write expressions for the relative momenta $q_1$ and 
$q_2$ in the $e^+e^-$ CM frame. In the quarkonium rest frame, 
$q_i$ is given by
%--------------
\begin{equation}
%--------------
q_i = |\bm{q}_i|
(0, \sin\theta_i\cos\phi_i, 
    \sin\theta_i\sin\phi_i, \cos \theta_i),
%--------------
\label{q_i}%
\end{equation}
%--------------
where $\theta_i$ and $\phi_i$ are the polar and azimuthal angles 
of $q_i$. Boosting Eq.~(\ref{q_i}) from the $P_i$ rest frame to
the $e^+e^-$ CM frame, one obtains
%--------------
\begin{subequations}
\begin{eqnarray}
%--------------
q_1^\ast&=&|\bm{q}_1|
      (+\gamma_1\beta_1 \cos\theta_1,
          \sin\theta_1\cos\phi_1,
          \sin\theta_1\sin\phi_1, 
       \gamma_1\cos\theta_1),
\\
q_2^\ast&=&|\bm{q}_2|
      (-\gamma_2\beta_2 \cos\theta_2,
          \sin\theta_2\cos\phi_2,
          \sin\theta_2\sin\phi_2, 
       \gamma_2\cos\theta_2),
%--------------
\end{eqnarray}
\end{subequations}
%--------------
where
%--------------
\begin{subequations}
\begin{eqnarray}
%--------------
\gamma_i &=& E_i/\sqrt{E_i^2-P_{\textrm{CM}}^2},
\\
\gamma_i \beta_i &=& P_{\textrm{CM}}/\sqrt{E_i^2-P_{\textrm{CM}}^2}.
%--------------
\end{eqnarray}
\label{gamma_i}%
\end{subequations}
%--------------
Note that $|\bm{q}_i|=\sqrt{-q_i^2}$ is the magnitude of the 
three-vector, not in the $e^+e^-$ CM frame, but in the $P_i$
rest frame.

It follows from Eq.~(\ref{cmframe}) and the analogue of Eq.~(\ref{me}) for 
$\mathcal{A}_{Q}^\mu({}^3S_1,P_1,\lambda;{}^1S_0,P_2)$ that 
%------------------
\begin{subequations}
\begin{eqnarray}
%------------------
\mathcal{A}_{Q}^\mu({}^3S_1,P_1,0;{}^1S_0,P_2)
&=&0,
\\
\mathcal{A}_{Q}^\mu({}^3S_1,P_1,\pm;{}^1S_0,P_2)
&=&
\pm 
A_QP_{\textrm{CM}}\sqrt{s}
\,\epsilon^{\ast \mu} (\pm).
%------------------
\end{eqnarray}
\label{A_Q}%
\end{subequations}
%------------------
It is efficient to determine $A_Q$ by carrying out the computation of the 
amplitude $\mathcal{A}_{Q}^\mu({}^3S_1,P_1,\lambda;{}^1S_0,P_2)$ for one 
value of $\mu$ and one value of $\lambda$ such that $\epsilon^{\ast \mu} 
(\lambda)$ is nonzero.

%%%%%%%%%%%%%%%%%%%%%%%%%%%%%%%%%%%%%%%%%%%%%%%%%%%%%%%%%%%%%%%%%%%%%%%%%%%%%%
\section{Cross section \label{sec:x-section}}
%%%%%%%%%%%%%%%%%%%%%%%%%%%%%%%%%%%%%%%%%%%%%%%%%%%%%%%%%%%%%%%%%%%%%%%%%%%%%%

Making use of Eq.~(\ref{me}) and the explicit choices of helicity 
states in Eq.~(\ref{cmframe}), we find that 
%------------------
\begin{subequations}
\begin{eqnarray}
%------------------
\mathcal{A}^\mu[J/\psi(0)+\eta_c]
&=&0,
\\
\mathcal{A}^\mu[J/\psi(\pm)+\eta_c]
&=&
\pm 
AP_{\textrm{CM}}\sqrt{s}
\,\epsilon^{\ast\mu} (\pm).
%------------------
\end{eqnarray}
\label{A}%
\end{subequations}
%------------------
Comparing Eq.~(\ref{A}) with Eq.~(\ref{A_Q}) and making use of the 
resummed NRQCD expansion in Eq.~(\ref{A-HH3}), we see that 
%------------------
\begin{equation}
A=\frac{1}{2N_c}
\langle J/\psi(\lambda)|\psi^\dagger 
\bm{\sigma}\cdot\bm{\epsilon}(\lambda)\chi|0\rangle
\langle \eta_c|\psi^\dagger\chi|0\rangle
\left.\left[\frac{\sqrt{2m_1}\sqrt{2m_2}}{2E(q_1)\, 2E(q_2)} 
A_Q\right]\right|_{\bm{q}_1^2=\langle \bm{q}^2\rangle_{J/\psi},\,
\bm{q}_2^2=\langle \bm{q}^2\rangle_{\eta_c}}.
\label{A-A_Q}
\end{equation}
%------------------

$\mathcal{M}(\lambda)$, the $S$-matrix element for the process 
$e^+e^-\to J/\psi + \eta_c$, is defined in Eq.~(\ref{em}).
By making use of Eqs.~(\ref{me}) and (\ref{A}), one can evaluate 
$\mathcal{M}(\lambda)$:
%------------------
\begin{subequations}
\begin{eqnarray}
%------------------
\mathcal{M}(0)&=&0,\\
\mathcal{M}(\pm)&=&
\pm AP_{\textrm{CM}}\sqrt{s}\,L\cdot\epsilon^\ast(\pm).
%------------------
\end{eqnarray}
\label{em-simple}%
\end{subequations}
%------------------
The squared helicity amplitudes, summed over the spin states 
$s^+=\pm1/2$ and  $s^-=\pm1/2$ of the $e^+$
and $e^-$, respectively, can be obtained by using 
Eqs.~(\ref{cmframe}) and (\ref{em-simple}):
%------------------
\begin{equation}
%------------------
 \sum_{s^\pm=\pm1/2}  
|L\cdot\epsilon^\ast(\pm)|^2
=
\frac{e_c^2e^4}{s^2}
\textrm{Tr}[/\!\!\!k_1/\!\!\!\epsilon(\pm)/\!\!\!k_2%
/\!\!\!\epsilon^\ast(\pm)]
=
\frac{e_c^2e^4}{s}(1+\cos^2\theta),
%------------------
\label{LL-ee}
\end{equation}
%------------------
which lead to
%------------------
\begin{subequations}
\begin{eqnarray}
%------------------
\sum_{s^\pm=\pm 1/2}
|\mathcal{M}(0)|^2
&=&0,\\
\sum_{s^\pm=\pm 1/2}
|\mathcal{M}(\pm)|^2
&=&
e_c^2 e^4 |A|^2P_{\textrm{CM}}^2
(1+\cos^2\theta).
%------------------
\end{eqnarray}
\label{em-sq}%
\end{subequations}
%------------------
Averaging the squared helicity amplitude (\ref{em-sq}) over the 
lepton spins, dividing by the flux $2s$, and integrating over the
two-body phase space, we obtain the total cross section for
$e^+e^-\to J/\psi+\eta_c$:
%------------------
\begin{eqnarray}
%------------------
\sigma[e^+e^-\to J/\psi+\eta_c]
&=&
\frac{1}{2s}
\times \frac{1}{4}
\times \Phi_2
\int_{-1}^1\frac{d\cos\theta}{2}
\,
\sum_{\lambda=\pm}
\,\,
\sum_{s^\pm=\pm 1/2}
|\mathcal{M}(\lambda)|^2
\nonumber\\
&=&
\frac{16\pi^2}{3s} e_c^2\alpha^2
|A|^2P^2_{\textrm{CM}}\Phi_2\nonumber\\
&=&\frac{4\pi^2}{3N_c^2 s} e_c^2\alpha^2
P^2_{\textrm{CM}}\Phi_2
\langle\mathcal{O}_1\rangle_{J/\psi}\langle\mathcal{O}_1\rangle_{\eta_c}
\nonumber\\
&&\times\left.\left[\frac{2m_1\,2m_2}{4E^2(q_1)\, 4E^2(q_2)} 
|A_Q|^2\right]\right|_{\bm{q}_1^2=\langle \bm{q}^2\rangle_{J/\psi},\,
\bm{q}_2^2=\langle \bm{q}^2\rangle_{\eta_c}},
%------------------
\label{sigma}
\end{eqnarray}
%------------------
where 
%------------------
\begin{subequations}
\begin{eqnarray}
\langle\mathcal{O}_1\rangle_{J/\psi}&=&
\left|
\langle J/\psi(\lambda)|\psi^\dagger 
\bm{\sigma}\cdot\bm{\epsilon}(\lambda)\chi|0\rangle
\right|^2,
\label{psi-me-defn}\\
\langle\mathcal{O}_1\rangle_{\eta_c}&=&
\left|
\langle \eta_c|\psi^\dagger
\chi|0\rangle\right|^2,
\label{etac-me-defn}
\end{eqnarray}
\label{me-defns}%
\end{subequations}
%------------------
and $\Phi_2$ is the two-body phase space
%------------------
\begin{eqnarray}
%------------------
\Phi_2
=
\frac{1}{8\pi s}\lambda^{1/2}(s,m_{J/\psi}^2,m_{\eta_c}^2).
%------------------
\label{Phi2}
\end{eqnarray}
%------------------
Note that we use the physical masses for the $J/\psi$ and the $\eta_c$
in the phase space~(\ref{Phi2}).

\section{VMD treatment of the photon-fragmentation amplitude
\label{sec:VMD}}
%--------------------------------------------------------------------
In the part of the amplitude that comes from the photon-fragmentation
diagrams of the type in Fig.~\ref{fig2}, we can reduce the theoretical
uncertainty by making use of the VMD method to calculate the
fragmentation of the $\gamma^*$ into $J/\psi$
(Ref.~\cite{Bodwin:2006yd}). In Ref.~\cite{Bodwin:2006yd}, the process
$\gamma^*\to J/\psi$ has been calculated using the VMD
method. Using Eq.~(3) of Ref.~\cite{Bodwin:2006yd}, we find
that the $\gamma^*\to J/\psi$ coupling is
%------------------
\begin{equation}
%------------------
g_{J/\psi}=\left(
\frac{3m_{J/\psi}^3}{4\pi\alpha^2}\Gamma[J/\psi\to \ell^+\ell^-]
\right)^{1/2}.
%------------------
\label{gv}%
\end{equation}
%------------------
In order to implement the VMD calculation we must make the following
substitutions in the NRQCD calculation of the photon-fragmentation
diagrams:
%------------------
\begin{subequations}
\begin{eqnarray}
%------------------
e_c\sqrt{2m_1\langle \mathcal{O}_1\rangle_{J/\psi} }
&\to&
g_{J/\psi},
\label{gv-conversion}%
\\
\frac{\textrm{Tr}\big\{
(\gamma_\mu \otimes \mathbbm{1}
)
[\Pi_3(p_1,\bar{p}_1,\lambda) \otimes \pi_1
]
                 \big\}
     }{\sqrt{2N_c} \, 2E(q_1)}
&\to&
\epsilon^*_\mu(\lambda),
\label{eps-conversion}%
%------------------
\end{eqnarray}
\label{frag-conversion}%
\end{subequations}
%------------------
where $g_{J/\psi}$ is defined in Eq.~(\ref{gv}).

\section{Choice of the quarkonium masses \label{sec:masses}}

We now specify our choices of the quarkonium masses in our
computation. In computing diagrams involving on-shell quarks, such as
those in Fig.~\ref{fig1}, it is generally necessary, in order to
maintain gauge invariance, to choose the quarkonium masses so as to
respect the on-shell condition. Hence, we generally must choose
$\tilde{m}_i=2E(q_i)=2\sqrt{m_c^2+\bm{q}_i^2}$ 
in Eq.~(\ref{PCM}) in working out the kinematics.

An exception to this is in the computation of the photon-fragmentation
diagrams of the type in Fig.~\ref{fig2}. In this case, if we make use of
the VMD method for calculating the amplitude, we can maintain gauge
invariance even if we take $\tilde{m}_1$ to be $m_{J/\psi}$,
the physical $J/\psi$ mass. We still must choose $\tilde{m}_2=2E(q_2)$
for the $\eta_c$ mass, however. Since one generally reduces theoretical
uncertainties by eliminating $2m_c$ in favor of $m_{J/\psi}$, we choose
$\tilde{m}_1=m_{J/\psi}$ in the VMD calculation of the
photon-fragmentation diagrams.

The factor $\sqrt{2m_1}\sqrt{2m_2}$ in Eq.~(\ref{A-HH3}) arises from the 
relativistic normalizations of the states. In this case, we 
choose $m_i=2E(q_i)$. It turns out that this 
choice leads to a near cancellation of the dependence on $m_c$ in the amplitude 
at leading order in $v$ (Ref.~\cite{Braaten:2002fi}). Thus, this choice 
reduces the theoretical uncertainties that arise from the uncertainty in 
$m_c$. 

As we have already noted, we use the physical quarkonium masses
in computing the phase space in Eq.~(\ref{Phi2}). At first sight, this
choice might appear to be inconsistent with the choice $m_i=2E(q_i)$ in
Eq.~(\ref{A-HH3}), since the factors $m_i$ in Eq.~(\ref{A-HH3}) arise
from the normalizations of the states, which also enter into the phase
space. The choices that we have made amount to multiplying the amplitude
by the factors $\sqrt{2E(q_1)/m_{J/\psi}}$ and $\sqrt{2E(q_2)/m_{\eta_c}}$.
At the level of precision in  $v$ to which we work, these factors are
equivalent to unity.

%%%%%%%%%%%%%%%%%%%%%%%%%%%%%%%%%%%%%%%%%%%%%%%%%%%%%%%%%%%%%%%%%%%%%%%%%%%%%%
\section{Interference with the NLO amplitude \label{sec:interference}}
%%%%%%%%%%%%%%%%%%%%%%%%%%%%%%%%%%%%%%%%%%%%%%%%%%%%%%%%%%%%%%%%%%%%%%%%%%%%%%

As we have mentioned, the corrections to $\sigma[e^+e^-\to J/\psi
+\eta_c]$ at NLO in $\alpha_s$ have been calculated in
Ref.~\cite{Zhang:2005ch}. Because the amplitude for the
relativistic corrections has the same phase as the amplitude at leading 
order in $v$, we can infer, from the results of Ref.~\cite{Zhang:2005ch}, 
the contribution to the cross section of the interference between the 
amplitude at NLO in $\alpha_s$ and the amplitude for the relativistic 
corrections. 

First, let us define some notation. When we discuss cross sections
$\sigma$ and reduced hadronic amplitudes $A$ [Eq.~(\ref{me})], a
subscript $0$ indicates that the quantity is computed at leading order
in $v$, a subscript $v$ indicates that the quantity is resummed to all
orders in $v$, a subscript NLO on $A$ indicates the contribution to $A$
at NLO in $\alpha_s$, and a subscript NLO on $\sigma$ indicates the sum
of the contributions  to $\sigma$ through NLO in $\alpha_s$. A
superscript QCD indicates that only QCD contributions to the hadronic
amplitude have been included. The absence of a superscript QCD indicates
that both QCD and pure QED contributions to the hadronic amplitude
have been included. Using this notation, we have
%------------------
\begin{subequations}
\begin{eqnarray}
%------------------
\sigma_0&=&\mathcal{N}|A_0|^2,
\\
\sigma_v&=&\mathcal{N}
|A_v|^2,
\\
\sigma_0^{\textrm{QCD}}&=&\mathcal{N}
|A_0^{\textrm{QCD}}|^2,
\\
\sigma^{\textrm{QCD}}_{0,\,\textrm{NLO}}&=&
\mathcal{N}
\Big[
|A^{\textrm{QCD}}_0|^2
+2\textrm{Re}\big(A^{\textrm{QCD}}_0
A^{\ast\textrm{QCD}}_{\textrm{0,\,NLO}}\big)
\Big],
\label{sigma-NLO}%
%------------------
\end{eqnarray}
\label{sigma_i}%
\end{subequations}
%------------------
where the normalization factor $\mathcal{N}$ is that of Eq.~(\ref{sigma}) and 
is defined by
%------------------
\begin{equation}
%------------------
\mathcal{N}
=\frac{16\pi^2}{3s} e_c^2\alpha^2
P^2_{\textrm{CM}}\Phi_2,
%------------------
\label{norm}%
\end{equation}
%------------------
with $P_{\textrm{CM}}$ defined in Eq.~(\ref{PCM}) and $\Phi_2$ defined in
Eq.~(\ref{Phi2}).

The quantity $\sigma^{\textrm{QCD}}_{0,\,\textrm{NLO}}$ is computed in
Ref.~\cite{Zhang:2005ch}. On the other hand, we wish to compute the 
quantity
%------------------
\begin{equation}
%------------------
\sigma_{\textrm{tot}}=
\mathcal{N}
\Big[
|A_v|^2
+2\textrm{Re}\big(A_v
A^{\ast\textrm{QCD}}_{0,\,\textrm{NLO}}\big)
\Big].
%------------------
\label{sigmatot0}
\end{equation}
%------------------
Using the fact that $A_v$ and $A_0^{\textrm{QCD}}$ have the same phase, 
we can write
\begin{eqnarray}
2\textrm{Re}\big(A_v
A^{\ast\textrm{QCD}}_{0,\,\textrm{NLO}}\big)&=&
\frac{A_v}{A^{\textrm{QCD}}_0}\,
2\textrm{Re}\big(A^{\textrm{QCD}}_0
A^{\ast\textrm{QCD}}_{\textrm{0,\,NLO}}\big)\nonumber\\
&=&
\frac{1}{\mathcal{N}}
\sqrt{\sigma_v}\,
\frac{\sigma^{\textrm{QCD}}_{0,\,\textrm{NLO}}-\sigma_0^{\textrm{QCD}}}
{\sqrt{\sigma_0^{\textrm{QCD}}}}.
\label{NLO-contrib}
\end{eqnarray}
Thus, $\sigma_{\textrm{tot}}$ can be expressed in terms of 
$\sigma_v$, $\sigma_0^{\textrm{QCD}}$,
and $\sigma^{\textrm{QCD}}_{0,\,\textrm{NLO}}$:
%------------------
\begin{equation}
%------------------
\sigma_{\textrm{tot}}
=
\sigma_v+
\sqrt{\sigma_v}\,
\frac{\sigma^{\textrm{QCD}}_{0,\,\textrm{NLO}}-\sigma^{\textrm{QCD}}_0}
     {\sqrt{\sigma^{\textrm{QCD}}_0}}.
%------------------
\label{sigma-tot}
\end{equation}
%------------------

%%%%%%%%%%%%%%%%%%%%%%%%%%%%%%%%%%%%%%%%%%%%%%%%%%%%%%%%%%%%%%%%%%%%%%%%%%%%%%

%%%%%%%%%%%%%%%%%%%%%%%%%%%%%%%%%%%%%%%%%%%%%%%%%%%%%%%%%%%%%%%%%%%%%%%%%%%%%%
\section{Results \label{sec:results}}
%%%%%%%%%%%%%%%%%%%%%%%%%%%%%%%%%%%%%%%%%%%%%%%%%%%%%%%%%%%%%%%%%%%%%%%%%%%%%%

In this section, we present our numerical results. 

We compute $A_Q$ in Eq.~(\ref{A_Q}) from the Feynman diagrams in
Figs.~\ref{fig1} and \ref{fig2}, making use of the spin and color
projectors, as described in Section~\ref{sec:gamma-Q4}. We then carry
out the projection onto the $S$-wave states by performing the
integration over the angles of $\bm{q}_1$ and $\bm{q}_2$ numerically, as
indicated in Eq.~(\ref{amp-QQ-angle}). We substitute
$A_Q$ into Eq.~(\ref{A-A_Q}) to obtain $A$ and substitute $A$ into
Eq.~(\ref{sigma}) to obtain the cross section $\sigma_v$, which includes
the resummed relativistic corrections. We compute $\sigma_0$,
the cross section at leading order in $v$, by setting $\bm{q}_1^2=\bm{q}_2^2=0$
in the expressions for $\sigma_v$.

In carrying out this calculation, we make use of the matrix elements 
$\langle \mathcal{O}_1 \rangle_{J/\psi}$ and 
$\langle \mathcal{O}_1 \rangle_{\eta_c}$
and the ratios of matrix elements $\langle\bm{q}^2\rangle_{J/\psi}$ and 
$\langle\bm{q}^2\rangle_{\eta_c}$ from Tables~I and III of 
Ref.~\cite{BCKLY}. In Ref.~\cite{BCKLY}, various uncertainties were 
associated with these quantities. The uncertainties are correlated to 
varying degrees among the quantities. We recount the uncertainties here. 

There are theoretical uncertainties in the values of $\langle
\bm{q}^2\rangle_{J/\psi}$ and $\langle \bm{q}^2\rangle_{\eta_c}$ that
arise from the fact that the leading-potential approximation that is
used in Ref.~\cite{BCKLY} is accurate only up to corrections of relative
order $v^2$. These uncertainties are denoted by $\Delta\langle
\bm{q}^2\rangle_{J/\psi}$ and $\Delta\langle \bm{q}^2\rangle_{\eta_c}$,
respectively. There are uncertainties that arise from the scale
uncertainties in $\alpha_s$ and from neglecting 
next-to-next-to-leading-order (NNLO) corrections to
the $J/\psi$ and $\eta_c$ electromagnetic widths. They are denoted by
$\Delta{\rm NNLO}_{J/\psi}$ and $\Delta{\rm NNLO}_{\eta_c}$, respectively.
There are also uncertainties that are associated with the heavy-quark
mass $m_c$, the string tension $\sigma$, and the uncertainties in the
experimental measurements of $\Gamma[J/\psi\to e^+e^-]$ and
$\Gamma[\eta_c\to \gamma\gamma]$. These uncertainties are denoted by
$\Delta m_c$, $\Delta\sigma$, $\Delta\Gamma_{J/\psi}$, and
$\Delta\Gamma_{\eta_c}$, respectively. Finally, there is an uncertainty
that is associated with the use of the heavy-quark spin symmetry to
combine the values of the $\eta_c$ matrix elements that were obtained
from $\Gamma[\eta_c\to \gamma\gamma]$ with those that were obtained from
$\Gamma[J/\psi\to e^+e^-]$. It is denoted by $\Delta v^2$. 

The uncertainty estimates in Ref.~\cite{BCKLY} make use of the
standard NRQCD power-counting (velocity-scaling) rules
\cite{Bodwin:1994jh}. Alternative power-counting rules have been
proposed \cite{Brambilla:1999xf,Pineda:2000sz,Fleming:2000ib}, and
the use of these rules would lead to estimates for $\langle
\bm{q}^2\rangle_{J/\psi}$ and $\langle \bm{q}^2\rangle_{\eta_c}$ that 
are of relative order unity. However, lattice
calculations~\cite{bks,Bodwin:2005gg,Bodwin:2004up,Koma:2007jq} support
the notion that the standard NRQCD power-counting rules give an upper
bound on the uncertainties. Therefore, we make use of the uncertainty
estimates of Ref.~\cite{BCKLY}. (See Ref.~\cite{BCKLY} for a more
detailed discussion of these issues.)

The calculation requires some additional inputs. We take
$\sqrt{s}=10.58$~GeV. In order to maintain consistency with the
calculation at NLO in $\alpha_s$ in Ref.~\cite{Zhang:2005ch}, we
take $m_c$ to be the one-loop pole mass. The specific numerical value
that we use is\footnote{The most recent compilation of the Particle Data
Group \cite{Yao:2006px} suggests that the uncertainty
in $m_c$ may be a factor of two smaller than the uncertainty that we
have used here. However, since it is not clear that the systematic
errors are well understood in the various determinations that enter into
that compilation, we make a conservative choice of error bars here.}
\begin{equation}
m_c=1.4\pm 0.2\hbox{ GeV}.
\end{equation}
This choice of numerical value corresponds to the one in
Ref.~\cite{BCKLY}, and so, in determining the uncertainties that arise
from the uncertainty in $m_c$, we are able to make use of the
dependences of the matrix elements and ratios of matrix elements on
$m_c$ that are computed in Ref.~\cite{BCKLY}. For the electronic width 
of the $J/\psi$, which enters into the calculation of the VMD coupling 
$g_{J/\psi}$ from Eq.~(\ref{gv}), we take
\begin{equation}
\Gamma[J/\psi\to e^+e^-]=5.55\pm 0.14\pm 0.02~\hbox{keV}.
\label{psi-width-exp}%
\end{equation}
For the strong and electromagnetic couplings we take\footnote{We
compute $\alpha_s$ and $\alpha$ at each scale by making use of the code
GLOBAL ANALYSIS OF PARTICLE PROPERTIES (GAPP)~\cite{Erler:1998sy}.}
\begin{subequations}
\begin{eqnarray}
\alpha_s(10.58/4\hbox{~GeV})&=&0.26,\\
\alpha_s(10.58/2 \hbox{~GeV})&=&0.21,\\
\alpha_s (10.58 \hbox{~GeV})&=&0.17, 
\end{eqnarray}
\label{alphas}%
\end{subequations}
and
\begin{subequations}
\begin{eqnarray}
\alpha(10.58/4\hbox{~GeV})&=&(132.9)^{-1},\\
\alpha(10.58/2\hbox{~GeV})&=&(131.9)^{-1},\\
\alpha(10.58 \hbox{~GeV})&=&(130.9)^{-1},\\
\alpha(m_{J/\psi})&=&(132.6)^{-1}.
\end{eqnarray}
\label{alpha}%
\end{subequations}
We determine the central value of the scale for each coupling from the
momentum transfer at the relevant vertex. Let us call the virtual
photon that connects the lepton and $c$-quark lines photon 1, the
non-fragmentation virtual photon that connects $c$-quark lines
in Fig.~\ref{fig1} photon 2, and the virtual photon in the fragmentation
diagrams in Fig.~\ref{fig2} photon 3. At virtual-gluon vertices and at
photon-2 vertices, we take the scale to be half the CM energy; at 
photon-1 vertices, we take the scale to be the CM energy; at photon-3 vertices,
we take the scale to be the $J/\psi$ mass. In calculating the VMD coupling 
$g_{J/\psi}$ from Eq.~(\ref{gv}), we also take the scale of the
virtual-photon vertices to be the $J/\psi$ mass.
These choices are consistent with those in Ref.~\cite{BCKLY}. We also
use $m_{J/\psi}=3.096916$~GeV and $m_{\eta_c}=2.9798$~GeV
\cite{Yao:2006px}.

In computing the cross section $\sigma_{\textrm{tot}}$, which includes 
relativistic corrections, corrections of NLO in $\alpha_s$, and the 
interference between them, we make use of Eq.~(\ref{sigma-tot}). We 
compute the quantity $(\sigma^{\textrm{QCD}}_{0,\,\textrm{NLO}}-
\sigma^{\textrm{QCD}}_0)/\sqrt{\sigma^{\textrm{QCD}}_0}$ by making use 
of results \cite{ZGC} that have been provided by the authors of 
Ref.~\cite{Zhang:2005ch}. These results are shown in Table~\ref{table1}.
%%%%%%%%%%%%%%%%%%%%%%%%%%%%%%%%%%%%%%%%%%%%%%%%%%%%%%%%%%%%%%%%%%%%%%%%%%%%%%
\begin{table}[t]
\caption{\label{table1}%
The cross sections in units of fb that were obtained by Zhang, Gao, and
Chao~\cite{Zhang:2005ch,ZGC}. 
The first row below the headings contains the cross sections for
central values of $m_c$ and $\mu$. Subsequent rows contain the cross
sections for the plus and minus variations of $m_c$ and $\mu$ with
respect to their uncertainties.
The strong coupling constant was taken to be
$\alpha_s^\textrm{ZGC}(10.6/4~\textrm{GeV})=0.273$,
$\alpha_s^\textrm{ZGC}(10.6/2~\textrm{GeV})=0.211$, and
$\alpha_s^\textrm{ZGC}(10.6~\textrm{GeV})=0.174$~(Ref.~\cite{ZGC}).
}
\begin{ruledtabular}
\begin{tabular}{lcc}
Case   &                      $\sigma_0^\textrm{ZGC}$&  
                              $\sigma^{\textrm{ZGC}}_{0,\,\textrm{NLO}}$\,\\
\hline
central            &    5.8& 12.6\\
$+\Delta m_c$  &    4.8&  9.7\\
$-\Delta m_c$  &    6.7& 15.7\\
$+\Delta\mu$   &    3.9&  8.9\\
$-\Delta\mu$   &    9.6& 19.6
\end{tabular}
\end{ruledtabular}
\end{table}
%%%%%%%%%%%%%%%%%%%%%%%%%%%%%%%%%%%%%%%%%%%%%%%%%%%%%%%%%%%%%%%%%%%%%%%%%%%%%%
Here we have introduced an additional uncertainty $\Delta \mu$, which
accounts for effects of the uncertainty in the renormalization scale in
the NLO calculation. This uncertainty is determined by varying the
renormalization scale by a factor of two above and below its central
value of $10.58/2\hbox{~GeV}$.
In making use of the results from the
authors of Ref.~\cite{Zhang:2005ch}, we rescale the values of
$\alpha_s$, $\alpha$, the matrix elements
$\langle\mathcal{O}_1\rangle_{J/\psi}$ and
$\langle\mathcal{O}_1\rangle_{\eta_c}$, and the phase space so that they
conform to our choices. The rescaling is carried out as follows:
%------------------
\begin{equation}
%------------------
\frac{\sigma^{\textrm{QCD}}_{0,\,\textrm{NLO}}-\sigma^{\textrm{QCD}}_0}
     {\sqrt{\sigma^{\textrm{QCD}}_0}}
     =
\sqrt{\rho}\,
\frac{\sigma^{\textrm{ZGC}}_{0,\,\textrm{NLO}}-\sigma^{\textrm{ZGC}}_0}
     {\sqrt{\sigma^{\textrm{ZGC}}_0}},
%------------------
\label{2rho}
\end{equation}
%------------------
where the superscript ZGC indicates the value that was given by the
authors of Ref.~\cite{Zhang:2005ch}. The scaling factor $\rho$ is
defined by
%------------------
\begin{equation}
%------------------
\rho
     =
\left(\frac{\alpha(10.58\hbox{~GeV})
     =1/130.9}{\alpha^{\textrm{ZGC}}=1/137}\right)^2
\left(\frac{\alpha_s(\mu)}
           {\alpha_s^{\textrm{ZGC}}(\mu)}\right)^2
\frac{\langle \mathcal{O}_1\rangle_{J/\psi}\langle\mathcal{O}_1\rangle_{\eta_c}}
           {\left(\langle \mathcal{O}_1\rangle_{J/\psi}^{\textrm{ZGC}}\right)^2}
\frac{\Phi_2}{\Phi_2^{\textrm{ZGC}}}.
%------------------
\label{rho}
\end{equation}
%------------------
The values for $\Phi_2^{\textrm{ZGC}}$ and 
$\langle \mathcal{O}_1\rangle_{J/\psi}^{\textrm{ZGC}}$ are given by
%------------------
\begin{subequations}
\begin{eqnarray}
%------------------
\Phi_2^{\textrm{ZGC}}
&=&
\frac{1}{8\pi}\times\sqrt{1-\left(\frac{4m_c}{\sqrt{s}}\right)^2},
\\
\langle \mathcal{O}_1\rangle_{J/\psi}^{\textrm{ZGC}}
&=&\frac{3}{2\pi}\times
0.978\textrm{\,GeV}^3
=
0.467\textrm{\,GeV}^3.
%------------------
\end{eqnarray}
\end{subequations}
%------------------

Our numerical results are shown in Table~\ref{table2}.
The first row below the
headings gives the central values of the matrix elements, ratios of
matrix elements, and the cross sections.
%%%%%%%%%%%%%%%%%%%%%%%%%%%%%%%%%%%%%%%%%%%%%%%%%%%%%%%%%%%%%%%%%%%%%%%%%%%%%%
\begin{table}[ht]
\caption{ The matrix elements $\langle \mathcal{O}_1 \rangle_{J/\psi}$
and $\langle \mathcal{O}_1 \rangle_{\eta_c}$ in units of
$\textrm{GeV}^3$, the ratios of matrix elements $\langle \bm{q}^2
\rangle_{J/\psi}$ and $\langle \bm{q}^2 \rangle_{\eta_c}$ in units of
$\textrm{GeV}^2$, and the cross sections $\sigma_0$, $\sigma_v$, and
$\sigma_\textrm{tot}$ in units of fb. The first row below the headings
contains central values for the matrix elements, the ratios, and the
cross sections. Subsequent rows contain the maximum and minimum values
of each of these quantities that are obtained by varying the input
parameters with respect to each uncertainty. 
\label{table2}
}
\begin{ruledtabular}
\begin{tabular}{lccccccc}
Case &
$\langle \mathcal{O}_1\rangle_{J/\psi} $&
$\langle\bm{q}^2 \rangle_{J/\psi}$&
$\langle \mathcal{O}_1 \rangle_{\eta_c}$ &
$\langle \bm{q}^2 \rangle_{\eta_c}$ &
$\sigma_0$ &
$\sigma_v$ &
$\sigma_\textrm{tot}$
\\
\hline
central&
0.440 & 0.441 & 0.437 & 0.442 & 6.4 & 9.3 & 17.6\\
$+\Delta \langle \bm{q}^2\rangle_{J/\psi}$&
0.450 & 0.573 & 0.437 & 0.442 & 6.5 & 9.8 & 18.4\\
$-\Delta \langle \bm{q}^2\rangle_{J/\psi}$&
0.430 & 0.308 & 0.437 & 0.442 & 6.3 & 8.8 & 16.7\\
$+\Delta m_c$&
0.433 & 0.443 & 0.470 & 0.430 & 6.0 & 7.6 & 13.9\\
$-\Delta m_c$&
0.451 & 0.437 & 0.413 & 0.450 & 6.9 &11.8 & 22.8\\
$+\Delta \sigma$&
0.443 & 0.482 & 0.444 & 0.482 & 6.6 & 9.7 & 18.3\\
$-\Delta \sigma$&
0.437 & 0.400 & 0.431 & 0.403 & 6.3 & 8.9 & 16.9\\
$+\Delta \,\textrm{NNLO}_{J/\psi}$&
0.504 & 0.419 & 0.473 & 0.429 & 7.9 &11.3 & 21.5\\
$-\Delta \,\textrm{NNLO}_{J/\psi}$&
0.387 & 0.459 & 0.408 & 0.452 & 5.3 & 7.8 & 14.6\\
$+\Delta \Gamma_{J/\psi}$&
0.451 & 0.437 & 0.443 & 0.440 & 6.7 & 9.6 & 18.2\\
$-\Delta \Gamma_{J/\psi}$&
0.429 & 0.444 & 0.431 & 0.444 & 6.2 & 9.0 & 16.9\\
$+\Delta v^2$&
0.440 & 0.441 & 0.511 & 0.417 & 7.5 &10.8 & 20.4\\
$-\Delta v^2$&
0.440 & 0.441 & 0.364 & 0.467 & 5.3 & 7.8 & 14.7\\
$+\Delta \langle \bm{q}^2\rangle_{\eta_c}$&
0.440 & 0.441 & 0.461 & 0.574 & 6.8 &10.2 & 19.1\\
$-\Delta \langle \bm{q}^2\rangle_{\eta_c}$&
0.440 & 0.441 & 0.414 & 0.309 & 6.1 & 8.4 & 16.1\\
$+\Delta \,\textrm{NNLO}_{\eta_c}$&
0.440 & 0.441 & 0.474 & 0.429 & 7.0 &10.0 & 19.0\\
$-\Delta \,\textrm{NNLO}_{\eta_c}$&
0.440 & 0.441 & 0.408 & 0.452 & 6.0 & 8.7 & 16.4\\
$+\Delta \Gamma_{\eta_c}$&
0.440 & 0.441 & 0.487 & 0.425 & 7.2 &10.3 & 19.5\\
$-\Delta \Gamma_{\eta_c}$&                 
0.440 & 0.441 & 0.385 & 0.460 & 5.6 & 8.2 & 15.5\\ 
$+\Delta \mu$&                 
0.440 & 0.441 & 0.437 & 0.442 & 4.4 & 6.3 & 12.3\\ 
$-\Delta \mu$&                 
0.440 & 0.441 & 0.437 & 0.442 & 9.5 &13.9 & 25.0 \\ 
\end{tabular}
\end{ruledtabular}
\end{table}
%%%%%%%%%%%%%%%%%%%%%%%%%%%%%%%%%%%%%%%%%%%%%%%%%%%%%%%%%%%%%%%%%%%%%%
Subsequent rows contain the maximum and minimum values of each of these
quantities that are obtained by varying the input parameters with
respect to each of the uncertainties that we have described. The matrix
elements and the ratios, as well as their variations with respect to
each uncertainty, are taken from Tables I and III of Ref.~\cite{BCKLY}.
The deviations from the central values, given in the same order as the 
rows in Table~\ref{table2}, are as follows:
%---------------------
\begin{subequations}
\begin{eqnarray}
%---------------------
\sigma_0 &=&
6.4^{+0.1+0.5+0.1+1.5+0.3+1.1+0.4+0.5+0.7}_{
       -0.1-0.5-0.1-1.1-0.2-1.1-0.3-0.4-0.8}~\textrm{fb}
=6.4^{ + 2.1}_{ - 1.9}~\textrm{fb},
\\
\sigma_v &=&
9.3^{+0.5+2.5+0.4+2.0+0.3+1.5+0.9+0.7+1.0}_{
     -0.5-1.7-0.4-1.5-0.3-1.5-0.9-0.6-1.1}~\textrm{fb}
=9.3^{+3.9 }_{ -3.2 }~\textrm{fb},
\\
\sigma_\textrm{tot} &=&
17.6^{+0.8+5.3+0.7+3.9+0.7+2.8+1.6+1.4+1.9}_{
      -0.9-3.7-0.7-3.0-0.7-2.9-1.5-1.1-2.0}~\textrm{fb}
=17.6^{+7.8 }_{ -6.3 }~\textrm{fb}.
%---------------------
\end{eqnarray}
\end{subequations}
%---------------------
In the result for $\sigma_\textrm{tot}$ above, we have not included the
uncertainty $\Delta\mu$ that arises from varying the renormalization
scale. That uncertainty is  ${}^{+7.4}_{-5.3}$~fb.
This is, perhaps, an overestimate of the uncertainty from uncalculated
corrections of higher order in $\alpha_s$ and $v^2$, since it assumes
that our choice of renormalization scale may be wrong by as much as a
factor of two. Alternatively, one could estimate the uncertainty that
arises from uncalculated corrections in the following way. One could
take for the uncertainty associated with uncalculated corrections of
NNLO in $\alpha_s$ to be the quantity 
$\Delta\hbox{NNLO}=\alpha_s(\sigma_\textrm{0,NLO}^\textrm{QCD}-
\sigma_\textrm{0}^\textrm{QCD})\approx
1.32 \hbox{~fb}$, and one could take for the uncertainty associated
with uncalculated corrections of NLO in $\alpha_s$ and NLO in $v^2$ the
quantity $\Delta\hbox{NLO-$v^2$}=
v^2(\sigma_\textrm{0,NLO}^\textrm{QCD}-\sigma_\textrm{0}^\textrm{QCD})
\approx 1.89\hbox{~fb}$.
(Uncertainties of relative order $v^4$ are already included in
$\Delta\langle\bm{q}^2\rangle_{J/\psi}$ and
$\Delta\langle\bm{q}^2\rangle_{\eta_c}$.) If we add the uncertainty
$\Delta\mu$ in quadrature with the other uncertainties, then we obtain
\begin{equation}
\sigma_\textrm{tot}=17.6^{+10.7}_{-8.3}\hbox{~fb}.
\label{sigma-delta-mu}
\end{equation}
On the other hand, if we add $\Delta\hbox{NNLO}$ and 
$\Delta\hbox{NLO-$v^2$}$ in quadrature with the other uncertainties, 
then we obtain
\begin{equation}
\sigma_\textrm{tot}=17.6^{+8.1}_{-6.7}\hbox{~fb}.
\label{sigma-delta-nnlo}
\end{equation}
In addition to the uncertainties that we have included in
Eqs.~(\ref{sigma-delta-mu}) and (\ref{sigma-delta-nnlo}), there are
uncertainties that are associated with the NRQCD factorization
formula. A rigorous proof of NRQCD factorization for $\sigma[e^+e^-\to
J/\psi +\eta_c]$ does not exist. However, it seems likely, on the basis
of existing work on proving NRQCD factorization for other production
processes \cite{Nayak:2005rt}, that the corrections to the factorization
formula are of order $m_H^2/(s/4)\approx 34\%$, where $m_H$ is the mass
of either of the heavy quarkonia.

The various contributions to $\sigma_\textrm{tot}$ are as follows. The
cross section at leading order in $\alpha_s$ and $v$, $\sigma_0$,
contributes about $6.4$~fb, of which about $1.0$~fb comes from the pure
QED corrections. The direct relativistic corrections that are
associated with the process $e^+e^-\to J/\psi + \eta_c$ contribute
about $2.9$~fb. The corrections of NLO in $\alpha_s$ contribute about
$6.9$~fb, including the interference with the pure QED contribution. 
The interference between the relativistic corrections and the
corrections of NLO in $\alpha_s$ contributes about $1.4$~fb.

We have examined our numerical calculation in the limits $\langle
\bm{q}^2\rangle_{J/\psi}\to 0$ and $\langle \bm{q}^2\rangle_{\eta_c}\to
0$ and find that it agrees with the analytic results in
Refs.~\cite{Braaten:2002fi} and \cite{He:2007te} for the order-$v^2$
corrections to $\sigma[e^+e^-\to J/\psi +\eta_c]$.

The direct relativistic corrections to the process $e^+e^-\to J/\psi
+\eta_c$ itself are modest in size. $\sigma_v$ is about $45\%$ larger
than $\sigma_0$, but $\sigma_0$ already contains an implicit
relativistic correction factor of $0.96$ that arises from the use of the
hadron masses, rather than $2m_c$, in the phase space. Hence, the
enhancement from the direct relativistic correction is about $40\%$.
If we use the hadron masses in the phase space and keep only the
order-$v^2$ relativistic corrections to the squared amplitude, then we
find that the direct relativistic corrections increase the cross section
by about $45\%$. Thus, we see that the effects of resummation are not
large, suggesting that the velocity expansion of NRQCD is converging
well in this case.

As we have mentioned previously, the resummed result contains all of the
corrections that are associated with the momentum-space $Q\bar Q$
quarkonium wave function in the leading-potential approximation, up to the
ultraviolet cutoff of the NRQCD matrix elements. Hence, the modest size
of the relativistic corrections supports the conclusion in
Ref.~\cite{Bodwin:2006dm} that the effects of the finite width of the
momentum-space $Q\bar Q$ wave function are not dramatic, once one excludes
contributions from the large-momentum tails of the wave function. Those
contributions are included in the NRQCD formalism in the corrections of
higher order in $\alpha_s$.

The use of the VMD method, rather than the NRQCD method, in 
calculating the fragmentation amplitude in the pure QED contribution, 
has a small effect on the central value of the cross section. The use of 
the VMD method shifts the central value of $\sigma_v$ down by about 
$3\%$.

\section{Comparison with previous calculations \label{sec:comparison}}

Let us now compare our results with some of those from previous 
calculations. 

As we have already mentioned, the contribution to $\sigma[e^+e^-\to
J/\psi +\eta_c]$ at leading order in $\alpha_s$ and $v$ was 
first calculated in Refs.~\cite{Braaten:2002fi} and \cite{Liu:2002wq}.
There are some differences in these results, owing to different choices of
input parameters and the inclusion of pure QED corrections in
Ref.~\cite{Braaten:2002fi}. Let us focus on Ref.~\cite{Braaten:2002fi},
since the calculation in that paper is closer to the present one in
terms of input parameters and the treatment of pure QED corrections. The
result in Ref.~\cite{Braaten:2002fi} is $\sigma[e^+e^-\to
J/\psi+\eta_c]=3.78\pm 1.26~\hbox{fb}$. This result should be compared
with our result for $\sigma_0$, which is about $70\%$ larger. This
difference arises essentially because we have used the values for
$\langle \mathcal{O}_1\rangle_{J/\psi} $ and $\langle
\mathcal{O}_1\rangle_{\eta_c}$ from Ref.~\cite{BCKLY} (see
Table~\ref{table2}), while the authors of Ref.~\cite{Braaten:2002fi}
have used $\langle \mathcal{O}_1\rangle_{J/\psi} =\langle
\mathcal{O}_1\rangle_{\eta_c}=0.335\hbox{~GeV}^3$. This substantial
difference in the values of the matrix elements arises largely from the
inclusion of relativistic corrections to the electromagnetic decay widths
of the $J/\psi$ and the $\eta_c$ in analyses of Ref.~\cite{BCKLY}. The
relativistic corrections increase the sizes of $\langle
\mathcal{O}_1\rangle_{J/\psi}$ and $\langle\mathcal{O}_1\rangle_{\eta_c}$ 
by about $31\%$, where we are comparing in both instances with the 
matrix element that is extracted from $\Gamma[J/\psi\to e^+e^-]$ in 
Ref.~\cite{Braaten:2002fi}. The changes in the values of the matrix elements
lead to a $72\%$ change in the cross section. Other small
differences in our calculation relative to that in
Ref.~\cite{Braaten:2002fi} arise from using the VMD method to calculate
the fragmentation contribution to the pure QED amplitude (about $-8\%$),
from the use of the physical masses for the $J/\psi$ and the $\eta_c$ in
the phase space (about $-4\%$), and from taking into account the effects
of the running of $\alpha$ (about $10\%$). The error bar in the result
of Ref.~\cite{Braaten:2002fi} takes into account only the uncertainty
$\Delta m_c$. It is more than twice the size of the $\Delta m_c$ error bar
in $\sigma_0$ in our calculation. The error bar in our calculation is
reduced because the $\Delta m_c$ uncertainty in the matrix elements in
Ref.~\cite{BCKLY} was reduced by replacing certain factors of $2m_c$
with $m_{J/\psi}$.

In Ref.~\cite{Bodwin:2006ke}, a result $\sigma_\textrm{tot}=17.5\pm
5.7$~fb is given. Our calculation contains a number of refinements in
comparison with that of Ref.~\cite{Bodwin:2006ke}. Among them are the
use of the improved results for the matrix elements in
Ref.~\cite{BCKLY}, the use of the VMD method to calculate the
fragmentation contribution to the pure QED amplitude, the inclusion of
the effects of the running of $\alpha$, and the precise calculation of
the interference between the relativistic corrections and the
corrections of NLO in $\alpha_s$, rather than the use of an overall
$K$~factor to account for the corrections of NLO in $\alpha_s$. The
effects of these refinements cancel almost exactly in the central value for
the cross section. The error bars in the result of
Ref.~\cite{Bodwin:2006ke} include only the uncertainties $\Delta m_c$,
$\Delta \langle \bm{q}^2\rangle_{J/\psi}$, and $\Delta \langle
\bm{q}^2\rangle_{\eta_c}$ and are, therefore, somewhat smaller than the
error bars that we report here.

We can also compare our results with those of Ref.~\cite{He:2007te}. In 
that work, the quantities $\langle \mathcal{O}_1\rangle_{J/\psi}$, $\langle
\mathcal{O}_1\rangle_{\eta_c}$, and 
$\langle\bm{q}^2\rangle_{J/\psi}\langle\mathcal{O}_1\rangle_{J/\psi}
=\langle\bm{q}^2\rangle_{\eta_c}\langle\mathcal{O}_1\rangle_{\eta_c}$
were determined by comparing theoretical expressions for
$\Gamma[J/\psi\to e^+e^-]$, $\Gamma[\eta_c\to\gamma\gamma]$, and
$\Gamma[J/\psi\to \textrm{light hadrons}]$ with the experimental
measurements of those widths. The resulting values are $\langle
\mathcal{O}_1\rangle_{J/\psi}^\textrm{HFC}=0.573\hbox{~GeV}^3$, $\langle
\mathcal{O}_1\rangle_{\eta_c}^\textrm{HFC}=0.432\hbox{~GeV}^3$,
$\langle\bm{q}^2\rangle_{J/\psi}^\textrm{HFC}=0.202\hbox{~GeV}^2$, and
$\langle\bm{q}^2\rangle_{\eta_c}^\textrm{HFC}=0.268\hbox{~GeV}^2$,
where the superscript HFC denotes the value that was given in 
Ref.~\cite{He:2007te}. The
value of $\langle \mathcal{O}_1\rangle_{J/\psi}$ is about $30\%$
larger than the one that we employ, and the value of $\langle
\mathcal{O}_1\rangle_{\eta_c}$ is about $1\%$ smaller. The values of
$\langle\bm{q}^2\rangle_{J/\psi}$ and $\langle \bm{q}^2
\rangle_{\eta_c}$ are smaller than the values that we use by about
$54\%$ and $39\%$, respectively, and are considerably smaller than 
expectations from the NRQCD velocity-scaling rules. 
As was discussed in Ref.~\cite{BCKLY}, the smaller values of 
$\langle\bm{q}^2\rangle_{J/\psi}$ and $\langle
\bm{q}^2 \rangle_{\eta_c}$ arise in Ref.~\cite{He:2007te} because the
theoretical expression for $\Gamma[J/\psi\to \textrm{light hadrons}]$
contains a very large relativistic correction. We regard this as an
indication that the velocity-expansion for that process is not under
control.

The authors of Ref.~\cite{He:2007te} find that the direct relativistic
corrections to $\sigma[e^+e^-\to J/\psi+\eta_c]$ enhance the cross
section by about $26\%$. If we ignore the effects of pure QED contributions 
and resummation, we find an enhancement from direct relativistic 
corrections of about $56\%$. The difference
presumably arises from the use of smaller values of
$\langle\bm{q}^2\rangle_{J/\psi}$ and $\langle \bm{q}^2
\rangle_{\eta_c}$ in Ref.~\cite{He:2007te}.

In Ref.~\cite{He:2007te}, the central value for the total cross section
is $\sigma_\textrm{tot}^\textrm{HFC}=20.04$~fb. This result does not
include the pure QED contribution, the contribution from the interference
between the corrections of NLO in $\alpha_s$ and the relativistic
corrections, and the effects of resummation. The corresponding quantity
in our calculation is $14.7$~fb. Thus, we see that the result of
Ref.~\cite{He:2007te} is $37\%$ larger than ours. The main sources of
this difference are the use of a larger value of $\langle
\mathcal{O}_1\rangle_{J/\psi}$, which would increase our cross
section by about $30\%$, the use of a larger value of the strong coupling
($\alpha_s^\textrm{HFC}=0.2592$), which would increase our cross
by about $47\%$,\footnote{In Ref.~\cite{He:2007te} the cross
section at NLO in $\alpha_s$ was computed using $\alpha_s=0.2592$ and
$\mu=3.00\hbox{~GeV}$ (Ref.~\cite{ZGC}). In order to find the effect of this
choice of $\alpha_s$ and $\mu$ on our calculation, we compute
$\sigma^\textrm{ZGC}_0$ and $\sigma^\textrm{ZGC}_{0,\,\textrm{NLO}}$ at
$\alpha_s=0.2592$ and $\mu=3.00$, using the cross sections in the last row
of Table~\ref{table1} as inputs. We then use these values for 
$\sigma^\textrm{ZGC}_0$ and $\sigma^\textrm{ZGC}_{0,\,\textrm{NLO}}$
to evaluate Eq.~(\ref{2rho}).} the use of a smaller value of the
electromagnetic coupling ($\alpha^\textrm{HFC}=1/137$), which would
decrease our cross section by about $9\%$, the use of a larger value of
the charm-quark mass ($m_c^\textrm{HFC}=1.5$~GeV), which would decrease
our cross section at fixed values of the NRQCD matrix elements by about
$12\%$, and the use of smaller values of
$\langle\bm{q}^2\rangle_{J/\psi}$ and $\langle \bm{q}^2
\rangle_{\eta_c}$, which would decrease our cross section by about
$9\%$.

In Ref.~\cite{He:2007te}, the dependence of the cross section on $m_c$
is given. As $m_c$ is varied from $1.4$~GeV to $1.6$~GeV, a change in 
the cross section of $+37\%$ is found. In contrast, for this variation in 
$m_c$, we find a change in the cross section of $-30\%$.
Presumably the difference arises because, in the method that is used to
determine $\langle \mathcal{O}_1\rangle_{J/\psi}$ and $\langle
\mathcal{O}_1\rangle_{\eta_c}$ in Ref.~\cite{He:2007te}, those
quantities are proportional to $m_c^2$. In the method that is used in
Ref.~\cite{BCKLY} to determine $\langle \mathcal{O}_1\rangle_{J/\psi}$ and
$\langle \mathcal{O}_1\rangle_{\eta_c}$, the dependence of those
quantities on $m_c$ is much milder, partly because some factors of
$2m_c$ are replaced with $m_{J/\psi}$ in the theoretical expressions. 
The authors of
Ref.~\cite{He:2007te} have not estimated the sizes of uncertainties that
arise from other sources, and so it is not clear whether their method of
calculation leads to a more precise prediction for the cross section
than the one that we have used.

%%%%%%%%%%%%%%%%%%%%%%%%%%%%%%%%%%%%%%%%%%%%%%%%%%%%%%%%%%%%%%%%%%%%%%%%%%%%%%
\section{Summary and Discussion \label{sec:discussion}}
%%%%%%%%%%%%%%%%%%%%%%%%%%%%%%%%%%%%%%%%%%%%%%%%%%%%%%%%%%%%%%%%%%%%%%%%%%%%%%

For a number of years, the discrepancy between theoretical predictions 
for the exclusive double-charmonium 
cross section $\sigma[e^+e^- \to J/\psi + \eta_c]$
and experimental measurements has posed
a significant challenge to our understanding of quarkonium production.
Changes in the measured values of the cross section have reduced the
discrepancy somewhat \cite{Abe:2004ww,Aubert:2005tj}. More recently,
calculations of the corrections of NLO in $\alpha_s$ \cite{Zhang:2005ch}
and relativistic corrections \cite{Bodwin:2006ke,He:2007te} have
increased the theoretical prediction for the cross section by almost an
order of magnitude. The shifts in the theoretical and experimental
central values for the cross section have resolved the outstanding
discrepancy. However, in the absence of an analysis of the theoretical
uncertainties, the meaning of the apparent agreement between theory and
experiment is unclear.

In this paper, we have carried out a new computation of the relativistic
corrections to $\sigma[e^+e^-\to J/\psi + \eta_c]$, with the goals of
adding certain refinements to the calculation and making a more precise
estimate of the theoretical uncertainties. Some of the refinements,
relative to the calculation of Ref.~\cite{Bodwin:2006ke}, are the use of
the VMD method to calculate the fragmentation contribution to the pure
QED amplitude, the inclusion of the effects of the running of $\alpha$,
and the inclusion of a precise calculation of the interference between
the relativistic corrections and the corrections of NLO in $\alpha_s$,
as opposed to the use of a simple $K$-factor estimate. A further
significant refinement in our calculation is the use of an improved
determination of the relevant NRQCD matrix elements at leading order in
$v^2$ and at NLO in $v^2$ (Ref.~\cite{BCKLY}). This determination
includes an analysis of the correlated uncertainties in the matrix
elements. Our calculation exploits this information to give a much
more complete estimate of the uncertainties than was given in
Ref.~\cite{Bodwin:2006ke}.

Our calculation differs from the one in Ref.~\cite{He:2007te} in
that we include pure QED corrections, we take into account the effects
of the running of $\alpha$, we include the interference between the
relativistic corrections and the corrections of NLO in $\alpha_s$, and
we resum a class of relativistic corrections. As we discuss in
Section~\ref{sec:comparison}, the calculation of Ref.~\cite{He:2007te}
makes use of matrix elements that differ significantly in numerical
value from those that we use. Ref.~\cite{He:2007te} includes a 
discussion of scale uncertainties and the effect of the uncertainty in 
$m_c$, but does not provide an overall error bar for the cross 
section.

In our calculation, the relativistic corrections to
$\sigma[e^+e^-\to J/\psi + \eta_c]$ arise from two sources. The first,
direct source consists of the relativistic corrections to the process
$e^+e^-\to J/\psi + \eta_c$ itself. These increase the cross section by
about $40\%$. The second, indirect source of
relativistic corrections derives from the relativistic corrections to
the electromagnetic decay widths of the $J/\psi$ and the $\eta_c$, which
enter into the matrix-element determinations of Ref.~\cite{BCKLY}. 
These corrections increase 
the cross section by about $88\%$ (Ref.~\cite{BCKLY}).
(Other, smaller
corrections result in a net change in $\langle \mathcal{O}_1\rangle_{J/\psi}
\langle \mathcal{O}_1 \rangle_{\eta_c}$
of $72\%$ relative to the value of $\langle \mathcal{O}_1\rangle_{J/\psi}
\langle \mathcal{O}_1 \rangle_{\eta_c}$ that was used in the 
calculation of Ref.~\cite{Braaten:2002fi}.)
The inclusion of corrections of NLO in $\alpha_s$ further increases the
cross section by about $89\%$, of which about $15\%$ comes from the
interference between the relativistic corrections and the corrections of
NLO in $\alpha_s$.

Our principal results are given in Eqs.~(\ref{sigma-delta-mu}) and
(\ref{sigma-delta-nnlo}). In the former result, the uncertainties that
arise from uncalculated higher-order corrections are estimated by
varying the renormalization scale. In the latter result, those
uncertainties are assumed to be given by their nominal sizes, namely,
$\alpha_s$ and $v^2$ times the contribution to the cross section of
NLO in $\alpha_s$. In addition, there are uncertainties that result
from the use of the NRQCD factorization formula for the cross section,
which we estimate to be about $34\%$.

The central value for $\sigma[e^+e^-\to J/\psi + \eta_c]$ that we 
obtain is essentially the same as that of Ref.~\cite{Bodwin:2006ke}.
The effects of the various refinements that we have mentioned largely cancel.
However, some of the refinements allow us to constrain the theoretical 
uncertainties more tightly. Because we have included more sources of 
uncertainty in our estimates, our error bars are significantly larger 
than those in Ref.~\cite{Bodwin:2006ke}.

Our results for $\sigma[e^+e^-\to J/\psi + \eta_c]$ also agree, within
uncertainties, with the result of Ref.~\cite{He:2007te}. To some extent,
the effects of our use of different values of the matrix elements and
other input parameters are canceled by our inclusion of additional 
corrections. In Ref.~\cite{He:2007te}, the dependence of the cross
section on $m_c$ is given. That dependence is similar in magnitude but
opposite in sign to the one that we find, presumably because the authors
of Ref.~\cite{He:2007te} use a method to determine the NRQCD matrix
elements that is quite different from the method in Ref.~\cite{BCKLY}.
The authors of Ref.~\cite{He:2007te} have not estimated other
uncertainties, and so it is not clear whether their method of
calculation yields a result that is more precise or less precise than
ours.

As we have mentioned, in our calculation, we resum a class of
relativistic corrections to all orders in $v$. These corrections include
all of the relativistic corrections that are contained in the
color-singlet $Q\bar Q$ quarkonium wave function, up to the ultraviolet
cutoff of the NRQCD matrix elements. The effect of the resummation
beyond relative order $v^2$ is small, indicating that the velocity
expansion converges well for this process. The fact that the direct
relativistic corrections are modest in size supports the conclusion in
Ref.~\cite{Bodwin:2006dm} that the effects of the finite width of the
momentum-space $Q\bar Q$ wave function are not dramatic, once one
excludes contributions from the large-momentum tails of the wave
function that are contained in corrections of higher order in $\alpha_s$.

Let us discuss the prospects for decreasing the
uncertainties in our calculation. The largest uncertainty arises from
the uncalculated terms of relative order $\alpha_s v^2$ and relative
order $\alpha_s^2$. This uncertainty may be as large as 
${}^{+42\%}_{-30\%}$. A complete calculation of the order-$\alpha_s v^2$
corrections, which are the larger ones, seems quite feasible. The
calculation of the corrections of order-$\alpha_s^2$ would be a major
undertaking, but is not out of the question. The next largest source of
uncertainty arises from the use of the NRQCD factorization formalism
itself, which may lead to an uncertainty of about $34\%$. 
A more thorough understanding of the issues that are involved in
constructing a rigorous proof of a factorization theorem for
$\sigma[e^+e^-\to J/\psi + \eta_c]$ may lead to a different estimate of
these uncertainties. It is also conceivable that one could prove a
``higher-twist'' factorization theorem that would allow one to carry out
a systematic computation of corrections to the existing NRQCD
factorization formula. The uncertainties that arise from the use of the
NRQCD factorization formalism presumably would decrease as the CM energy
of the process $e^+e^-\to J/\psi + \eta_c$ increases. However, there are
no prospects for measuring $\sigma[e^+e^-\to J/\psi + \eta_c]$ at higher
energies in the immediate future. The uncertainty in $m_c$ is the next
most important source of theoretical uncertainty. We estimate the
resulting uncertainty in the cross section to be ${}^{+30\%}_{-21\%}$.
We can expect to see some progress in reducing this uncertainty,
particularly from lattice determinations of $m_c$.

Our result for $\sigma[e^+e^-\to J/\psi + \eta_c]$ agrees,
within errors, with the measurements of the Belle and BABAR experiments.
The uncertainties in our result are quite large, and, of course, it
would be desirable to reduce these uncertainties, so as to sharpen this
test of the NRQCD factorization approach to quarkonium production.
Nevertheless, it seems fair to conclude that the long-standing
discrepancy between the theoretical prediction for $\sigma[e^+e^-\to
J/\psi + \eta_c]$ and the experimental measurements has been resolved.

%--------------------------------------------------------------------
\begin{acknowledgments}
We thank Kuang-Ta Chao for providing us with numerical values of the
NLO cross section for various values of the input parameters. We 
are grateful to Jens Erler for providing us with the latest 
version of the code GAPP and for explaining its use.
We also thank Taewon Kim for checking the central value of the
cross section.
JL thanks the High Energy Physics Theory Group at Argonne National 
Laboratory for its hospitality while this work was carried out.
Work in the High Energy Physics Division at Argonne
National Laboratory is supported by the U.~S.~Department of Energy,
Division of High Energy Physics, under Contract No.~DE-AC02-06CH11357.
The work of JL was supported by the Korea Research Foundation
under MOEHRD Basic Research Promotion grant KRF-2004-015-C00092
and by a Korea University Grant.
The work of CY was supported by the Korea Research Foundation
under MOEHRD Basic Research Promotion grant KRF-2006-311-C00020.
\end{acknowledgments}

%%  APPENDIX

%\appendix

%--------------------------------------------------------------------
%\section{}

\end{document}